\documentclass[preprint,3p]{elsarticle}
\pdfoutput=1
\usepackage[toc,page]{appendix}
\usepackage{graphicx}
\usepackage{multicol}
\usepackage{psfrag}
\usepackage{amssymb}
\usepackage{amsthm}
\usepackage{lineno}
\usepackage{amsfonts}
\usepackage{amsmath}
\usepackage{caption}
\usepackage{bm}
\usepackage{framed}
\usepackage{color}
\usepackage{hyperref}
\usepackage[english]{algorithm2e} 
\usepackage[normalem]{ulem}
\hypersetup{colorlinks,breaklinks,
           linkcolor=blue,urlcolor=blue,
           anchorcolor=blue,citecolor=blue}
\usepackage[caption=false]{subfig}
\usepackage{float}
\usepackage{comment}

\newtheorem{lemma}{Lemma}[section]

\newtheorem*{remark}{Remark}
\graphicspath{{fig/}}

\renewcommand{\c}{\mathbf{c}}

\providecommand{\q}{\mathbf{q}}

\journal{Elsevier}
\begin{document}
\begin{frontmatter}
\title{A Low-rank Control Variate for Multilevel Monte Carlo Simulation of High-dimensional Uncertain Systems}
\author[label1]{Hillary Fairbanks}
\author[label2]{Alireza Doostan\corref{cor1}}
\ead{alireza.doostan@colorado.edu}
\cortext[cor1]{Corresponding Author: Alireza Doostan}
\author[label3]{Christian Ketelsen}
\author[label4]{Gianluca Iaccarino}

\address[label1]{Applied Mathematics and Statistics Department, University of Colorado, Boulder, CO 80309, USA}
\address[label2]{Aerospace Engineering Sciences Department, University of Colorado, Boulder, CO 80309, USA}
\address[label3]{Computer Science Department, University of Colorado, Boulder, CO 80309, USA}
\address[label4]{Mechanical Engineering Department, Stanford University, Stanford, CA 94305, USA}

\begin{abstract}

Multilevel Monte Carlo (MLMC) is a recently proposed variation of Monte Carlo (MC) simulation that achieves variance reduction by simulating the governing equations on a series of spatial (or temporal) grids with increasing resolution. Instead of directly employing the fine grid solutions, MLMC estimates the expectation of the quantity of interest from the coarsest grid solutions as well as differences between each two consecutive grid solutions. When the differences corresponding to finer grids become smaller, hence less variable, fewer MC realizations of finer grid solutions are needed to compute the difference expectations, thus leading to a reduction in the overall work. This paper presents an extension of MLMC, {\color{black} referred to as} {\it multilevel control variates (MLCV)}, where a low-rank approximation to the solution on each grid, obtained primarily based on coarser grid solutions, is used as a control variate for estimating the expectations involved in MLMC. Cost estimates as well as numerical examples are presented to demonstrate the advantage of this new MLCV approach over the standard MLMC when the solution of interest {\color{black}admits a low-rank approximation and the cost of simulating finer grids grows fast}. 
\end{abstract}


\end{frontmatter}

\section{Introduction}

The use of uncertainty quantification as a tool to {\color{black}assess} the prediction accuracy of simulation models of physical systems has been increasing at a rapid rate over the last decade. By accounting for the uncertainties of input data in models, such as initial conditions, boundary conditions, or other physical parameters, {\color{black}the objective is to establish the predictive capabilities of simulations by}
quantifying the uncertainty in the quantities of interest (QoI's). To this end and within the probabilistic framework, several methods, e.g., polynomial chaos expansions \cite{Ghanem03,Xiu02,Doostan11a} and stochastic collocation \cite{Xiu05,Mathelin03}, have been developed and proven successful in various applications. However, it is known that the computational cost of these methods grows rapidly as a function of the number of random variables describing model uncertainties, a phenomenon referred to as {\it curse of dimensionality}.

An alternative class of techniques rely on the Monte Carlo (MC) simulation or its variants, where the statistics of the QoI are estimated using an ensemble of (random) realizations of the QoI. The cost of such estimations, while may be prohibitive, is {\it formally} independent of the number of input variables. In details, let $\bm\xi = (\xi_1,\dots,\xi_d)$ denote the $d$-vector of random variables, with joint probability density function $\rho_{\bm\xi}(\bm\xi)$, representing the uncertainty in the inputs. Let $Q=Q(\bm\xi)$ denote a scalar-valued QoI depending on $\bm\xi$ and $Q_M$ its approximation obtained via simulation. The subscript $M$ denotes the number of {\color{black}deterministic} degrees of freedom, e.g., number of grid points in a finite element model, controlling the accuracy of $Q_M$ relative to $Q$. The goal is to approximate the statistics of $Q$, e.g., the mean of $Q$, $\mathbb{E}[Q]$, using the realizations of $Q_M$. Given a set of $N$ samples of inputs, each denoted by $\bm\xi^{(i)}$ and drawn according to $\rho_{\bm\xi}(\bm\xi)$, and the corresponding realizations of $Q_M$, given by $Q_M^{(i)} = Q_M(\bm\xi^{(i)})$, the MC approximation of $\mathbb{E}[Q] $ is
\begin{equation}\label{eq:mc_estimator}
\mathbb{E}[Q] \approx \mathbb{E}[Q_M] \approx \hat{Q}_{M,N}^{MC} = \frac{1}{N}\sum _{i=1}^N  Q_M^{(i)}.
\end{equation}
Following the notation in \cite{Cliffe11}, $\hat{Q}_{M,N}^{MC}$ in (\ref{eq:mc_estimator}) is the MC estimator of $\mathbb{E}[Q_M]$ using $N$ samples of $Q_M$ with the Mean Square Error (MSE)
\begin{equation}\label{eq:mcmse}
\text{MSE}(\hat{Q}_{M,N}^{MC},\mathbb{E}[Q]) = \frac{1}{N} \mathbb{V}[Q_M]+ \left(\mathbb{E}[Q_M-Q]\right)^2, 
\end{equation}
where $\mathbb{V}$ is the variance operator and $\text{MSE}(\hat{Q}_{M,N}^{MC},\mathbb{E}[Q])$ denotes the MSE of $\hat{Q}_{M,N}^{MC}$ with respect to $\mathbb{E}[Q]$. We note that, in this paper, the hat operator indicates the MC estimator of the corresponding expectation. In (\ref{eq:mcmse}), the MSE is decomposed into the {\it sampling error} $\frac{1}{N} \mathbb{V}[Q_M]$, controlled by the variance of $Q_M$ and the number of samples, and the {\it discretization error} $\left(\mathbb{E}[Q_M-Q]\right)^2$, which measures how closely the model simulates the true solution. As can be seen from (\ref{eq:mcmse}), the sampling error decays slowly as a function of $N$, but with a rate that is independent of the dimension $d$, implying that the standard MC simulation does not formally suffer from the curse of dimensionality. 

Aside from the necessary refinement in the model to reduce the discretization error, there are only two options to improve the MSE of a MC estimate: increasing the sample size $N$ or using a variance reduction technique. Due to the cost incurred by the first option, it is more practical to consider the use of a variance reduction technique, such as importance sampling or control variates (CV) \cite{Asmussen07}. In particular, the CV approach considers a second quantity $Z$, such that it is correlated with $Q_M$, is cheaper than $Q_M$ to evaluate, and whose expectation is either known or can be approximated with relatively small cost. Then a new variable,
\[
W = Q_M - \theta(Z - \mathbb{E}[Z]),
\]
is constructed that has the same mean as $Q_M$, i.e., $\mathbb{E}[W]= \mathbb{E}[Q_M]$, thus suggesting the use of MC estimate of $\mathbb{E}[W]$ as a proxy for $\mathbb{E}[Q_M]$.  {\color{black}In doing so, the gain is that, depending on the choice of $\theta$, the MC estimator of $\mathbb{E}[W]$ features a reduced MSE (or variance). Stated differently, a smaller number of $W$ realizations, hence $Q_M$ realizations, are needed for a comparable MSE when CV is applied. For scenarios when $Z$ is poorly correlated with $Q_M$, a notable MSE reduction is not observed. If, in addition, the cost of estimating $\mathbb{E}[Z]$ is large, it is likely that this CV will not result in a cost improvement over standard MC.}

Multilevel Monte Carlo (MLMC), proposed in \cite{Heinrich01,Giles08}, is a generalization of CV, which constructs a sequence of control variates $Z$ based on approximations of $Q$ on a set of models that are cheaper to simulate than the one for $Q_M$, hence the term {\it multilevel}. A common example of a cheaper model is to approximate $Q$ on coarser grids with number of degrees of freedom smaller than $M$. While the notion of levels can go beyond a grid-based construction, we limit the scope of this study to such an approach. For the interest of a simpler introduction, we delay the full presentation of MLMC to Section \ref{sec:background}, and instead focus on the two-level formulation next. 

Taking $\theta=1$ and $Z = Q_m$, with $m<M$, to be the QoI approximated from a coarser grid than that of $Q_M$, the two-level MLMC variable is given by $W = Q_M - (Q_m-\mathbb{E}[Q_m]) = \mathbb{E}[Q_m] + (Q_M - Q_m)$, with expected value 
\begin{align}
\label{eq:2_level_mlmc}
\mathbb{E}[W] &= \mathbb{E}[Q_m] + \mathbb{E}[Q_M - Q_m] \\
		       &=\mathbb{E}[Q_M].\nonumber
\end{align}
To approximate $\mathbb{E}[W]$, or equivalently $\mathbb{E}[Q_M]$, MC is applied independently to the two expectations in the right-hand-side of the first equation in (\ref{eq:2_level_mlmc}), 
\begin{equation}
\label{eq:2_level_mlmc_est}
\hat{W} = \frac{1}{N_m} \sum\limits_{i=1}^{N_m}Q_m^{(i)}+ \frac{1}{N_M} \sum\limits_{i=1}^{N_M}(Q_M^{(i)} - Q_m^{(i)}).
\end{equation}
As compared to the standard MC estimator of $\mathbb{E}[Q_M]$ given in (\ref{eq:mc_estimator}), the estimation of $\mathbb{E}[Q_m]$ in (\ref{eq:2_level_mlmc_est}) also involves drawing samples of $Q_m$, which are less expensive. More importantly, when $Q_m$ is close to $Q_M$, estimating $\mathbb{E}[Q_M - Q_m]$ requires fewer samples of $Q_M$, as $(Q_M - Q_m)$ features a smaller variance. In practice, depending on the cost of simulating the two models as well as the variances of  $Q_m$ and $(Q_M - Q_m)$, the numbers of samples of $Q_m$, $N_m$,  and $Q_M$, $N_M$, are selected such that the overall estimation cost, for a given accuracy, is minimal. MLMC expands upon this concept by including multiple levels, as delineated in Section \ref{sec:background}. 

Very recently the combination of MLMC and CV, {\color{black}referred to as {\it multilevel control variates}} (MLCV), has been used with the aim of further reducing the variance of $(Q_M - Q_m)$ and thereby improving the computational cost~\cite{Nobile15}. As an approach to MLCV, in \cite{Nobile15}, the authors consider the solution to a linear diffusion problem, where the diffusion coefficient is modeled by a rough random field represented by a large number of independent random variables. An auxiliary diffusion problem with smoothed (low-dimensional) coefficient is developed, whose solution -- computed via stochastic collocation -- is employed as the control variate for the MLMC simulation of the original problem. This particular construction of the control variate relies on the accuracy of the smoothed problem in predicting the solution to the original one.  Other related work \cite{Speight09,Vidal15} establishes multilevel control variates within a MC simulation framework. For instance, in \cite{Vidal15}, a model reduction approach based on the reduced basis method, see, e.g., \cite{Rozza08}, is considered for the case of linear, stochastic elliptic PDEs. In particular, in an off-line stage, a set of basis functions, i.e., reduced basis, is identified from the realizations of the solution on a target mesh. The PDE solution is then approximated in an increasing number of reduced basis functions, via a discontinuous Galerkin formulation, which are then employed as control variates. The levels here are defined based on the size of the reduced basis used, as opposed to the number of grid points in the spatial discretization of the PDE.  

With the aim of reducing the cost of MLMC, this work proposes a different construction of control variates that may be applicable to more general classes of problems. More precisely, we create a control variate $Z$ for $(Q_M-Q_m)$ that is obtained based on a low-rank approximation of $Q_M$ using a {\it small} set of selected samples of fine grid solution identified from the realizations of the coarse grid solution. This low-rank approximation of the fine grid solution, inspired by the work in \cite{Narayan14,Zhu14,Doostan07,Doostan16}, consists of three main steps. In the first step, we identify a reduced basis, say of size $r\ll M$, for the coarse grid solution, together with an interpolation rule that gives an arbitrary realization of the coarse grid solution in that basis. For this purpose, we borrow ideas from matrix interpolative decomposition (MID) as presented in \cite{Cheng05}. The second step entails the identification of a reduced basis for the fine grid solution. For this, we generate the fine counterpart of the coarse grid reduced basis, which requires $r$ realizations of the fine grid solution. In the third step, we apply the same interpolation rule, as for the coarse grid solution, to generate the low-rank approximation of its fine counterpart (and subsequently $Q_M$). Once $Z$ is formulated, the control variate equation 
\begin{equation}\label{eq:intro_cv_term}
W = (Q_M - Q_m) - \theta(Z - \mathbb{E}[Z]),\nonumber
\end{equation}
is used to estimate $\mathbb{E}[Q_M- Q_m]$, which requires setting $\theta$ and estimating $\mathbb{E}[Z]$. For these, we consider work from \cite{Pasupathy12}, where $\theta$ is chosen such that the MSE of $\hat{W}$, with respect to $\mathbb{E}[Q_M- Q_m]$, is minimized. In addition, $\mathbb{E}[Z]$ is estimated primarily from an independent MC simulation of the coarse grid solution. 

An important feature of this MLCV approach is worthwhile highlighting. Depending on the rank of the fine model solution and the cost of simulating $Q_m$, relative to $Q_M$, $Z$ may be sampled with a cost that is considerably smaller than that of $Q_M$. This, together with the MSE reduction achieved by $\hat{W}$, will lead to a smaller number of samples of $Q_M$ required for a similar accuracy as in standard MLMC. 

This paper is organized into four sections. A background on MLMC is provided in Section \ref{sec:background}, including the setup and improvements it provides over MC, as well as a discussion of MLMC theory. Section \ref{sec:MLCV} focuses on the formulation of the MLCV method of this paper, including the construction of the control variate and the MLCV estimator such that the MSE is minimized. This section also includes an algorithm to outline the process, as well as a brief discussion regarding considerations to take into account when taking this MLCV approach. In Section \ref{sec:results} the numerical results of this new method as well as comparison with MLMC are presented using two test problems. The first test considers an eigenvalue problem associated with a linear elasticity problem in an L-shaped domain, followed by the second case, a thermally driven flow problem in a square domain.

\section{Background on Multilevel Monte Carlo}
\label{sec:background}

From \cite{Heinrich01,Giles08}, we next describe the use of MLMC as a variance reduction technique for the MC simulation of differential equations with random data. Consider a sequence of spatial discretizations, i.e., grids, for the governing equations with increasing accuracy, where each discretization is indexed by a {\it level} parameter $\ell$, $\ell=0,\dots,L$. Let $M_{\ell}$ denote the number of degrees of freedom in the level $\ell$ discretization model, such that $M_0 <M_1< \hdots <M_L$. Unlike in MC, MLMC does not directly approximate the expectation of QoI on the finest (target) level $L$. Instead, it uses the telescoping sum 
\begin{equation}\label{eq:T_Sum}
\mathbb{E}[Q]\approx \mathbb{E}[Q_L] = \mathbb{E}[Q_0] + \sum\limits_{\ell=1}^L  \mathbb{E}[Q_\ell - Q_{\ell-1}],
\end{equation}
which is the expected value of the QoI on the coarse level plus the sum of expectations of correction terms. Here, $Q_\ell$ is the QoI determined from the level $\ell$ grid with $M_\ell$ degrees of freedom. Defining the multilevel correction variable 
\begin{equation}\label{eq:YL}
Y_{\ell} = Q_\ell - Q_{\ell-1},
\end{equation}
for $\ell=1,\hdots, L $, and $Y_0=Q_0$, the sum in (\ref{eq:T_Sum}) can be rewritten as 
\[
\mathbb{E}[Q_L] = \sum\limits_{\ell=0}^L  \mathbb{E}[Y_{\ell}].
\]
In MLMC, the expectation of each correction variable $Y_{\ell}$ is independently computed with $N_\ell$ MC samples of $Y_{\ell}$, 
\begin{equation}\label{eq:Yhat}
 \mathbb{E}[Y_{\ell}] \approx \hat{Y}_\ell = \frac{1}{N_{\ell}} \sum\limits_{i=1}^{N_{\ell}} Y^{(i)}_{\ell} = \frac{1}{N_{\ell}} \sum\limits_{i=1}^{N_{\ell}} \left(Q^{(i)}_\ell - Q^{(i)}_{\ell-1}\right),
 \end{equation}
where each $Q^{(i)}_\ell$ and $Q^{(i)}_{\ell-1}$ is generated by applying the same realization ${\bm \xi ^{(i)}}$ to the level $\ell$ and $\ell-1$ grids, respectively. Building on the notation once more, the MLMC estimator, $\hat{Q}_{L}^{ML}$, is defined to be
\[
\hat{Q}^{ML}_L = \sum\limits_{\ell=0}^L  \hat{Y}_{\ell}.
\]
When $Q_\ell$ converges to $Q$ as a function of $\ell$, the variance of $Y_\ell$ converges to zero. This allows for $\hat{Y}_\ell$ to converge with fewer samples of $Q_\ell$ for larger, and more expensive, levels $\ell$. To determine how to pick $N_\ell$ such that the same level of accuracy as in MC is reached with a reduced computational cost, the MSE of $\hat{Q}_{L}^{ML}$ must first be considered.  

\subsection{MSE of MLMC}

The MSE of the MLMC estimator, much like that of the MC estimator in (\ref{eq:mcmse}), can be decomposed into the sampling and discretization errors, given by
\begin{equation}\label{eq:mlmcmse}
\text{MSE}(\hat{Q}_{L}^{ML},\mathbb{E}[Q])=\sum\limits_{\ell=0}^L \frac{1}{N_{\ell}}  \mathbb{V}[Y_{\ell}] + \left(\mathbb{E}[Q_L-Q]\right)^2,
\end{equation}
where the sampling error of MLMC is a sum of MC sampling errors on each level. For a desired MSE bound of magnitude $\varepsilon ^2$, the sampling error (as well as the discretization error) is ideally required to be bounded by $\varepsilon ^2/2$; that is,
\begin{equation}\label{eq:mlmc_sampling_error}
\sum\limits_{\ell=0}^L \frac{1}{N_{\ell}} \mathbb{V}[Y_{\ell}] \le \frac{\varepsilon ^2}{2}.
\end{equation}
Provided an estimate of  $\mathbb{V}[Y_{\ell}]$, (\ref{eq:mlmc_sampling_error}) serves as a constraint to estimate the number of samples $N_\ell$ drawn from each level.

\subsection{Number of Samples $N_\ell$ for $\hat{Y}_\ell$}

In MLMC, $N_\ell$ is determined such that the total cost, given by
\begin{equation}\label{eq:mlmc_totalcost}
\mathcal{C}(\hat{Q}_{L}^{ML}) = \sum\limits_{\ell=0}^L  N_{\ell} \mathcal{C}_{\ell},
\end{equation}
 is minimized, where $\mathcal{C}_{\ell}$ is the cost of generating a sample of $Y_\ell$ and is given by 
\begin{equation}
\label{eq:cost_Y_l}
\mathcal{C}_{\ell}=\mathcal{C}(Y_{\ell})=\mathcal{C}(Q_\ell)+\mathcal{C}(Q_{\ell-1}),
\end{equation}
for $\ell = 1,\hdots , L$, and $\mathcal{C}_{0} =\mathcal{C}(Q_0)$. In practice, the average CPU time for obtaining a sample of $Y_\ell$ can be used to determine $\mathcal{C}_\ell$. Alternatively, when the computational complexity of the adopted solvers with respect to $M_\ell$ is known, a relation of the form 
\[
\mathcal{C}_{\ell} \lesssim M_{\ell}^{\gamma},\quad \gamma >0,
\]
may be considered, where $\lesssim$ denotes inequality up to a positive constant. By using the method of Lagrange multipliers to minimize the total cost in (\ref{eq:mlmc_totalcost}), with an $\varepsilon^2/2$ sampling error constraint in (\ref{eq:mlmc_sampling_error}), the ideal number of samples $N_\ell$ from each level can be computed from 
\begin{equation}\label{eq:nl}
N_{\ell} \ge \frac{2}{\varepsilon ^2} \left[ \ \sum \limits_{k=0}^L \sqrt{ \mathbb{V}[Y_k] \mathcal{C}_{k} }\right] \sqrt{\frac{\mathbb{V}[Y_{\ell}]}{\mathcal{C}_{\ell}}},
\end{equation}
see, e.g., \cite{Giles13}. In order to calculate the values of $N_{\ell}$, a {\it pilot run} of MLMC must first be implemented; that is, MLMC is completed on a relatively small sample size that may be constant across all levels. By doing so, the {\color{black}expectation,} variance, and cost on each level can be estimated to determine {\color{black}the required number of levels and optimal} $N_\ell$ from (\ref{eq:nl}), {\color{black} such that a given MSE tolerance can be achieved}. Subsequently, a second run of MLMC is completed, this time using only the calculated sample size $N_\ell$ for each level. To make the method as efficient as possible, simulations from the pilot run are incorporated into the second run of MLMC.
\subsection{MLMC Convergence Guarantee}
\label{subsec:mlmc_theory}

In order to guarantee an $\varepsilon ^2$ bound on the MSE in (\ref{eq:mlmcmse}), we turn to the MLMC theorem as presented in \cite{Cliffe11}. While the number of samples on each level can be determined such that the desired sampling error bound $\varepsilon ^2/2$ is attained, it does not imply the same for the discretization error. The MLMC theorem provides the constraints in order to maintain the prescribed MSE bound. Assuming that there exists an integer $s>1$ such that $M_{\ell} = sM_{\ell -1}$, for $\ell =1,\hdots , L$, and 
\begin{enumerate}
\item $|\mathbb{E}[Q_\ell- Q]|\lesssim M_\ell^{-\alpha}$,
\item $\mathbb{V}[Y_\ell]\lesssim M_\ell^{-\beta}$,
\item $\mathcal{C}_\ell \lesssim M_\ell^\gamma$,
\end{enumerate}
for some constants $\gamma,\beta>0$ and $\alpha \geq \frac{1}{2}\text{min}(\beta , \gamma)$, Theorem 1 in \cite{Cliffe11} states that there exists a positive integer $L$ depending on $\varepsilon$ and a sequence $\{ N_\ell \}_{\ell=0}^L$, such that $\text{MSE}(\hat{Q}_{L}^{ML},\mathbb{E}[Q])<\varepsilon ^2$.
This relation indicates whether or not it is necessary to refine the model further to meet the desired discretization error. In Section \ref{sec:results}, estimates of $\alpha$ and $\beta$ will be provided as an illustration that the theory holds for the numerical experiments of this study.

\section{Multilevel Control Variates}
\label{sec:MLCV}

It has been shown that MLMC is more cost effective than MC for many stochastic differential equations \cite{Heinrich01,Giles08,Cliffe11,Teckentrup13,Butler15}. Due to the success of this method, we consider continuing along the path of variance -- more precisely MSE -- reduction by applying control variates to each $\mathbb{E}[Y_\ell]$ estimate with the aim of reducing the number of samples $N_\ell$ required by MLMC or the required work to achieve a desired MSE $\varepsilon^2$. In detail, given the multilevel correction variable $Y_{\ell}$, as defined in (\ref{eq:YL}), and a correlated variable $Z_{\ell}$, a general MLCV correction variable is defined as
\begin{equation}
\label{eqn:control_eq_mlcv}
W_{\ell} = Y_{\ell} - \theta_{\ell}(Z_{\ell} - \mathbb{E}[Z_{\ell}]),
\end{equation}
for $\ell = 1,\hdots , L$, and $W_0 = Y_0$. If $\mathbb{E}[Z_{\ell}]$ is known, it follows that $\mathbb{E}[W_{\ell}]=\mathbb{E}[Y_{\ell}]$, and $\theta_{\ell}$ can be set such that the MSE of the estimator of $\mathbb{E}[W_\ell]$, $\hat{W}_\ell$, is optimized. However, it is often the case that $\mathbb{E}[Z_{\ell}]$ is unknown and thus must be estimated. In the following subsections we will discuss our formulation of the control variate $Z_\ell$ and the approximation of $\mathbb{E}[Z_{\ell}]$. The discussion will include the decomposition of the MSE of $\hat{W}_\ell$ with respect to $\mathbb{E}[Y_{\ell}]$ and the optimal number of samples needed to reach a desired sampling error. 

\subsection{$Z_{\ell}$ Formulation}
\label{subsec:z_mid}

We consider a formulation of $Z_{\ell}$ based on a low-rank approximation of $Y_{\ell}$ in (\ref{eq:YL}). Specifically,
\begin{equation}\label{eq:Z}
Z_{\ell} = Q^{ID}_{\ell} - Q_{\ell-1}, 
\end{equation}
where $Q_{\ell-1}$ is {\color{black}as defined in (\ref{eq:YL})}, and $Q^{ID}_{\ell}$ is a low-rank approximation of $Q_\ell$ based on samples of $Q_{\ell-1}$ and a pre-determined basis for level $\ell$ solution described next. {\color{black}Let $Q_\ell$ depend on a solution-dependent, vector-valued quantity $\q_\ell(\bm\xi)\in \mathbb{R}^{m_\ell}$; that is, 
\[
Q_\ell = Q\left(\q _\ell(\bm\xi)\right). 
\]
For instance, the (spatial) mean heat flux along a boundary, $Q_\ell$, can be computed based on the values of the heat flux, $\q_\ell$, on the boundary grid nodes. }

To construct $Q^{ID}_{\ell}$, we assume the approximation to $\q$ on level $\ell$, $\q_\ell$, admits an accurate representation in a reduced basis of (small) cardinality $r\ll m_\ell$. Inspired by~\cite{Narayan14,Zhu14,Doostan07,Doostan16}, such a low-rank approximation of $\q_\ell$ may be achieved by identifying a reduced basis for $\q_{\ell-1}$ along with an approximation rule representing an arbitrary sample of $\q_{\ell-1}$ in that basis. {\color{black}In details, some $N_\ell\ge m_\ell$
realizations of $\q_{\ell-1}$ corresponding to random samples $\{ {\bm \xi}^{(i)}\}_{i=1}^{N_{\ell}}$, are generated and organized in an $m_{\ell-1}\times N_{\ell}$ {\it coarse grid data matrix}}
\[
\bm{U}_{\ell-1}: =\left[ \q _{\ell -1}\left({\bm \xi ^{(1)}}\right) \ \ \q _{\ell -1}\left({\bm \xi ^{(2)}}\right) \ \cdots \ \q _{\ell -1}\left({\bm \xi ^{(N_{\ell})}}\right) \right].
\]
{\color{black}We note that, here, $N_\ell$ is different from the sample size $N_\ell$ in MLMC and will be defined more precisely in Section \ref{subsubsec:pilot}.}
To find a reduced basis for $\q_{\ell-1}$, we consider a rank $r$ factorization of $\bm{U}_{\ell-1}$ using a subset of its columns. While a number of tools are available for this purpose \cite{Elhamifar09,Mahoney09,Halko11,Dyer15}, we employ the so-called interpolative decomposition (ID) of $\bm{U}_{\ell-1}$ \cite{Gu96,Cheng05,Halko11}, 
\begin{equation}\label{eq:CGmid} 
\bm{U}_{\ell-1} \approx \bm{U}^c_{\ell-1} \bm{C}_{\ell-1},
\end{equation}
where the $m_{\ell-1}\times r$ {\it column skeleton} matrix
\[
\bm{U}^c_{\ell-1} =
\left[ \q _{\ell -1}\left({\bm \xi ^{(i_1)}}\right) \ \ \q _{\ell -1}\left({\bm \xi ^{(i_2)}}\right) \ \cdots \ \q _{\ell -1}\left({\bm \xi ^{(i_r)}}\right) \right]
\]
consists of $r$ columns of $\bm{U}_{\ell-1}$ identified via pivoted rank-revealing QR factorization of $\bm{U}_{\ell-1}$, and $\bm{C}_{\ell-1}$ is an $r \times N_{\ell}$ {\it coefficient matrix} as specified in Appendix~A. Stated differently, (\ref{eq:CGmid}) gives a rank $r$ approximation of $\q_{\ell-1}(\bm\xi^{(i)})$ in a {\it reduced basis} $\bm{U}^c_{\ell-1}$ consisting of $r$ realizations of $\q_{\ell-1}$. For $\q_{\ell-1}(\bm \xi)$ evaluated at an arbitrary $\bm\xi$, a rank $r$ representation of the form
\begin{equation}
\label{eqn:rank_r_q} 
\q_{\ell-1}(\bm \xi)\approx \bm{U}^c_{\ell-1}\ \mathbf{c}_{\ell-1}(\bm\xi)
\end{equation}
can be generated by computing the $r$-vector of coefficients $\mathbf{c}_{\ell-1}(\bm\xi)$ via least squares approximation. Specifically,
\begin{equation}
\label{eqn:least_squares} 
\mathbf{c}_{\ell-1}(\bm\xi) = \arg\min\limits_{\hat{\mathbf{c}}\in \mathbb{R}^r} \Vert \bm{U}^c_{\ell-1}\ \hat{\mathbf{c}} -\q_{\ell-1}(\bm \xi) \Vert _2,
\end{equation}
{\color{black}
which can be computed by via SVD or QR decomposition. Further discussion related to the complexity of solving for the coefficients, $\mathbf{c}_{\ell-1}(\bm\xi)$, can be found in Section \ref{subsec:mlcv_cost}.}

\begin{remark}
\label{rem:rank_r}
It is worthwhile highlighting that, in practice, $r$ is not known {\it a priori} and has to be chosen such that the approximation in (\ref{eq:CGmid}) achieves a desired accuracy. Additionally, to reveal the rank $r$ of $\bm U_{\ell-1}$ from (\ref{eq:CGmid}), $m_{\ell-1}$ must satisfy the condition $m_{\ell-1}\ge r$, as $r\le\min\{m_{\ell-1},N_\ell\}$. 
\end{remark}

\begin{remark}
\label{rem:rank_r}
The rank $r$ may vary from one discretization level to another; however, for the interest of a simpler notation, we suppress the dependence of $r$ on $\ell$. 
\end{remark}

From this process, we identify samples $\{ \bm{\xi}^{(i_k)} \}_{k=1}^{r}$ using which the corresponding {\it fine grid} reduced basis and approximation can be determined. Specifically, following \cite{Narayan14} and using the coarse grid coefficient vector $\mathbf{c}_{\ell-1}$ in (\ref{eqn:rank_r_q}), we now define the rank $r$ approximation of $\q_{\ell}$ as
\begin{equation}\label{eq:FGmid}
\q_{\ell}^{ID}(\bm \xi)= \bm{U}^c_{\ell}\ \mathbf{c}_{\ell-1}(\bm\xi),
\end{equation}
where the reduced basis
 $$\bm{U}^c_{\ell} =
\left[ \q _{\ell }\left({\bm \xi ^{(i_1)}}\right) \ \ \q _{\ell }\left({\bm \xi ^{(i_2)}}\right) \ \cdots \ \q _{\ell }\left({\bm \xi ^{(i_r)}}\right) \right]$$
is the $m_{\ell}\times N_{\ell}$ fine grid counterpart of $\bm{U}^c_{\ell-1}$. Finally, (\ref{eq:Z}) is fully specified by setting
\[
Q^{ID}_{\ell}=Q(\q^{ID}_{\ell}(\bm\xi) ).
\]
For a discussion on the convergence of the coarse-fine approximation $\q_{\ell}^{ID}$ in (\ref{eq:FGmid}), we refer the interested reader to Sections 3 and 4 of \cite{Narayan14}, where an identical construction is presented.  

\subsection{$\mathbb{E}[Z_{\ell}]$ Estimation}
\label{sec:EZ_estimate}

The numerical construction of $Z_\ell$ in (\ref{eq:Z}) does not lead to an analytic value for $\mathbb{E}[Z_{\ell}]$, as desired in standard control variate approaches. However, approximations to $\mathbb{E}[Z_{\ell}]$, denoted hereafter by $\bar{Z}_\ell$, may be alternatively used as long as they are {\color{black}not expensive} to generate~\cite{Schmeiser01,Emsermann02, Pasupathy12}. While there are multiple techniques to estimate $\mathbb{E}[Z_{\ell}]$, an independent MC simulation is used {\color{black}here}. In detail, let $N'_{\ell}$ be the number of MC samples collected from level $\ell-1$ to estimate $\mathbb{E}[Z_{\ell}]$. We define the MC estimator of $\mathbb{E}[Z_{\ell}]$ to be
\begin{equation}
\label{eq:z_bar}
\bar{Z}_{\ell} =\frac{1}{N'_{\ell}} \sum_{i=1}^{N'_{\ell}}Z_{\ell}^{(i)} = \frac{1}{N'_{\ell}} \sum_{i=1}^{N'_{\ell}}\left( Q^{ID (i)}_{\ell} - Q^{(i)}_{\ell-1} \right),
\end{equation}
where $Q^{ID (i)}_{\ell}$ is generated following the procedure of Section \ref{subsec:z_mid}. An advantage to this MC method is that the estimate $\bar{Z}_{\ell}$ only relies on additional level $\ell-1$ samples without any need to draw additional level $\ell$ samples. Thus, given the reduced bases $\bm{U}^c_{\ell-1}$ and $\bm{U}^c_{\ell}$, the primary cost of finding $\bar{Z}_{\ell}$ is running the level $\ell-1$ model to obtain $N'_{\ell}$ coarse grid samples, which is given by 
\begin{equation}\nonumber
\mathcal{C}(\bar{Z}_{\ell}) = N'_{\ell}\ \mathcal{C}( Q^{ID}_{\ell}) + N'_{\ell}\ \mathcal{C}( Q_{\ell-1}) \approx N'_{\ell}\ \mathcal{C}(Q_{\ell-1}).
 \end{equation}

 A more detailed quantification of $\mathcal{C}( Q^{ID}_{\ell})$ will be provided in Section \ref{subsec:mlcv_cost}. Next, we focus on the MSE reduction achieved with this method of control variates, and subsequently, the estimation of $\tilde{N}_\ell$ and $N'_\ell$.

\subsection{MSE Reduction via $W_\ell$}

Now that we have defined $Z_{\ell}$ and $\bar{Z}_{\ell}$ we can update (\ref{eqn:control_eq_mlcv}) to get
\begin{equation}
\label{eqn:control_mlcv_zbar}
W_{\ell} = Y_{\ell} - \theta_{\ell}(Z_{\ell} - \bar{Z}_{\ell}),
\end{equation}
for $\ell = 1,\hdots , L$, and $W_0 = Y_0$. Given a fixed (independently computed) $\bar{Z}_{\ell}$, $\hat{W}_{\ell}$ is a biased estimator of $\mathbb{E}[Y_{\ell}]$ with the bias $\theta_\ell \delta_\ell$, where $\delta_\ell = \bar{Z}_{\ell} - \mathbb{E}[Z_{\ell}]$. However, $\delta_\ell$ may be thought of as a random variable with mean zero and variance $\mathbb{V}[Z_{\ell}]/N'_{\ell}$. {\color{black}Define $\tilde{N}_\ell$ to be the number of samples of $(Y_\ell, Z_\ell)$ from (\ref{eqn:control_mlcv_zbar}) and recall $N'_\ell$ is the number of samples used to calculate $\bar{Z}_\ell$.} Based on the work in \cite{Pasupathy12}, $\theta_{\ell}$ can be selected such that the MSE of $\hat{W}_{\ell}$ -- also averaged over possible realizations of $\delta_\ell$ -- is minimized. 

By doing so, it follows that, see \cite{Pasupathy12},
\begin{equation}\label{mseR}
\text{MSE}(\hat{W}_{\ell},\mathbb{E}[Y_\ell]) = \text{MSE}(\hat{Y}_{\ell},\mathbb{E}[Y_\ell]) \left[ 1-\rho_\ell^2 \left( \frac{1}{1+\tilde{N}_{\ell}/N'_{\ell}}\right)\right],
\end{equation} 
where 
\begin{equation}\label{r2}
\rho_\ell^2 = \frac{\left(\text{cov}(Y_{\ell},Z_{\ell})\right)^2}{\mathbb{V}[Z_{\ell}]\mathbb{V}[Y_{\ell}]} 
\end{equation}
and
\begin{equation}\label{beta}
  \theta^{*}_{\ell} = \frac{\text{cov}(Y_{\ell},Z_{\ell})}{\mathbb{V}[Z_{\ell}]}\left( \frac{1}{1+\tilde{N}_{\ell}/N'_{\ell}}\right). 
  \end{equation}
Since {\color{black}$0\leq\rho_\ell^2\leq 1$}, (\ref{mseR}) and (\ref{r2}) indicate that as the correlation between $Y_{\ell}$ and $Z_{\ell}$ increases, $\rho_\ell^2$ approaches one, and thus the {\it MSE reduction factor} $ \text{MSE}(\hat{W}_{\ell},\mathbb{E}[Y_\ell]) / \text{MSE}(\hat{Y}_{\ell},\mathbb{E}[Y_\ell])$ goes to zero. 
Updating $\hat{W}_\ell$ to account for the optimal MSE, the MLCV correction estimator is given by
\begin{equation}\label{cvem}
\hat{W}_{\ell} = \frac{1}{\tilde{N}_\ell}\sum\limits_{i=1}^{\tilde{N}_\ell}\left(Y^{(i)}_{\ell} - \theta_{\ell}^{*}(Z^{(i)}_{\ell} - \bar{Z}_{\ell})\right),
\end{equation}
for $\ell = 1,\hdots , L$, and $\hat{W}_0 = \hat{Y}_0$. Then the full MLCV estimator is defined as
\[
\hat{Q}^{MLCV}_L = \sum\limits_{\ell=0}^L  \hat{W}_{\ell}.
\]

In order to complete the task of calculating the number of samples $\tilde{N}_\ell$ per level to optimize the cost, we must find the MSE of this MLCV estimator first. It can be shown that 
\begin{equation}\nonumber
\text{MSE}(\hat{Q}^{MLCV}_L,\mathbb{E}[Q]) = \text{MSE} \left( \hat{Q}^{MLCV}_L, \mathbb{E}[Q_L] \right)  +\left(\mathbb{E}[Q_L - Q]\right)^2,
\end{equation}
where the MSE's are also averaged over the realizations of $\delta_\ell$. Expanding the sampling error term, we find
\begin{align*}
\text{MSE}(\hat{Q}^{MLCV}_L,\mathbb{E}[Q_L]) &=\mathbb{E}\left[ \left( \hat{Q}^{MLCV}_L- \mathbb{E}[Q_L] \right)^2\right] \\
&=  \mathbb{E}\left[ \left( \hat{W}_0 - \mathbb{E}[Y_0]\right)^2\right]+\cdots + \mathbb{E}\left[\left(\hat{W}_L - \mathbb{E}[Y_L]\right)^2\right] \\
&=  \sum\limits_{\ell=0}^L \text{MSE}(\hat{W}_\ell,\mathbb{E}[Y_\ell])  \\
&=  \sum\limits_{\ell=0}^L \text{MSE}(\hat{Y}_\ell,\mathbb{E}[Y_\ell]) \left[ 1-\rho_\ell^2 \left( \frac{1}{1+\tilde{N}_{\ell}/N'_{\ell}}\right)\right],
\end{align*}
where the last line is determined from (\ref{mseR}). Thus the MSE of $\hat{Q}^{MLCV}_L$ is given by
\begin{equation}\label{MLCVmse}
\text{MSE}(\hat{Q}^{MLCV}_L,\mathbb{E}[Q]) =  \sum\limits_{\ell=0}^L \frac{1}{\tilde{N}_{\ell}}\mathbb{V}[Y_\ell] \left[ 1-\rho_\ell^2 \left( \frac{1}{1+\tilde{N}_{\ell}/N'_{\ell}}\right)\right] + \left(\mathbb{E}[Q_L -Q]\right)^2,
\end{equation}
where we used the relation $\text{MSE}(\hat{Y}_\ell,\mathbb{E}[Y_\ell]) = \mathbb{V}[Y_\ell]/\tilde{N}_{\ell}$. Since the discretization error term in (\ref{MLCVmse}) is the same as in (\ref{eq:mlmcmse}), the MLMC convergence theory, as discussed in Section \ref{subsec:mlmc_theory}, can be applied to the simulations of {\color{black}$W_\ell$ in MLCV in order to guarantee a bound on the MSE.}

\subsection{Number of Samples $\tilde{N}_\ell$ for $\hat{W}_{\ell}$}\label{subsec:nlcv}

In MLMC the number of samples needed for each level is estimated in order to attain a bound on the sampling error component of the MSE, see (\ref{eq:nl}). In this formulation of MLCV, we must not only account for the sample size $\tilde{N}_{\ell}$, but also for $N^{'}_\ell$. The sample size $\tilde{N}_{\ell}$ is the required number of samples of $(Y_\ell, Z_\ell)$ to achieve a sampling error bound of $\varepsilon ^2/2$, while the value $N'_\ell$ is the number of samples used to find $\bar{Z}_{\ell}$. 

In practice, we are able to determine the ratio $\tilde{N}_{\ell}/N'_{\ell}$ from a number of pilot runs, as discussed in Section \ref{subsec:alg}, and update the MSE in (\ref{MLCVmse}) accordingly. Applying the same method to derive (\ref{eq:nl}) and fixing the ratio $\tilde{N}_{\ell}/N'_{\ell}$, the optimal number of samples $\tilde{N}_{\ell}$ for $\hat{W}_\ell$ can be determined from (\ref{MLCVmse}) and is given by
\begin{equation}\label{nlcv}
\tilde{N}_{\ell} = \frac{2}{\varepsilon ^2} \left[ \ \sum \limits_{k=0}^L \Big( \mathbb{V}[Y_k] \left[ 1-\rho^2_k \left( \frac{1}{1+\tilde{N}_k/N'_k }\right)\right]\mathcal{C}(W_k) \Big)^{1/2}\right] \sqrt{\frac{\mathbb{V}[Y_{\ell}]\left[ 1-\rho^2_\ell \left( \frac{1}{1+\tilde{N}_{\ell}/N'_{\ell} }\right)\right]}{\mathcal{C}(W_{\ell})}}, 
\end{equation}
where $\rho^2_k$ is also estimated from the pilot simulations. We note that in (\ref{nlcv}), we take
\begin{equation}
\mathcal{C}(W_{\ell})= \mathcal{C}_{\ell}. \nonumber
\end{equation}
This cost accounts for the samples of $Y_\ell$ and $Z_\ell$.  We consider the data acquisition for $\bar{Z}_{\ell}$ to be offline, thus the related cost is not included in the estimation of $\tilde{N}_{\ell}$; however, it is included in the total cost of implementing MLCV, as will be discussed in Section \ref{subsec:mlcv_cost}.   

\subsection{Number of Samples $N'_\ell$ for $\bar{Z}_{\ell}$}\label{subsec:npstar}

As can be observed from (\ref{MLCVmse}) or (\ref{nlcv}), the ratio $\tilde{N}_{\ell}/N'_{\ell}$ plays an important role in the MSE reduction achieved by $\hat{W}_\ell$ and the number of samples $\tilde{N}_{\ell}$ required by MLCV on each level. The smaller $\tilde{N}_{\ell}/N'_{\ell}$, the larger the MSE reduction {\color{black} and, hence, the smaller the estimated $\tilde{N}_{\ell}$ from (\ref{nlcv})}. However, a small ratio $\tilde{N}_{\ell}/N'_{\ell}$ requires a large number  $N'_{\ell}$ of independent samples to compute $\bar{Z}_\ell$ and, hence, a large overall computational cost. Therefore, it is necessary to determine an {\it optimal} value for $\tilde{N}_{\ell}/N'_{\ell}$.

In \cite{Pasupathy12}, the authors discuss the use of the so-called {\it generalized} MSE to find the optimal number of samples to minimize the product of the cost and $\text{MSE}(\hat{W}_\ell,\mathbb{E}[Y_\ell])$ from (\ref{mseR}). Let $\zeta = \mathcal{C}(Q_{\ell-1})/\mathcal{C}_\ell$ denote the cost per sample of $Z_\ell$ (the cost of a  level $\ell-1$ sample) divided by the cost per sample of $Y_\ell$, where $\mathcal{C}_{\ell}$ is given in (\ref{eq:cost_Y_l}). Following \cite{Pasupathy12} and by minimizing the product of the total cost required on each level -- given by $\zeta N'_{\ell} + (1+\zeta)\tilde{N}_{\ell}$ -- and $\text{MSE}(\hat{W}_\ell,\mathbb{E}[Y_\ell])$, the optimal ratio $\tilde{N}_{\ell}/N'_{\ell}$ is such that 
\begin{equation}\label{npstar}
N'_{\ell} = \text{max} \left( 0,s_1 -1  \right) \tilde{N}_{\ell}
\end{equation}
with
$$s_1 =\left[ \frac{\rho_\ell^2}{\zeta(1-\rho_\ell^2)} \right] ^{1/2}.$$
If $N'_{\ell} =0$ this implies we do not implement MLCV on level $\ell$, instead, we let $W_{\ell} = Y_{\ell}$. 

The potential issue with using (\ref{npstar}) is that for $\rho_\ell^2\approx 1$, the value of $N'_{\ell}$ can get quite large. In some tests this may occur, and depending on the desired implementation, the number of samples may alternatively be determined by
\begin{equation}\label{npstar2}
N'_{\ell}=s_2 \tilde{N}_{\ell},
\end{equation}
where the integer $s_2>1$ is preselected. In the numerical experiments of this work, we choose the minimum $N'_{\ell}$ from (\ref{npstar}) and (\ref{npstar2}), i.e.,
\begin{equation}\label{npstar3}
N'_{\ell} = \min\left(  s_2, \text{max} \left( 0,s_1 -1  \right)\right)\tilde{N}_{\ell},
\end{equation}
and we set $s_2=10$.

\subsection{MLCV Implementation Details and Algorithms}\label{subsec:alg}

In this section we discuss the implementation of MLCV as outlined in Algorithms \ref{alg:pilot} and \ref{alg:mlcv}. The MLCV algorithm can be organized into two different components: the pilot run and the full MLCV run. For both, we will describe the case that $\ell>0$. When $\ell=0$, we simply simulate the level $\ell=0$ model and take a sample average to calculate $\hat{W}_0 = \hat{Y}_0$. Additionally, we assume that $L$ is large enough to meet the desired discretization error. In practice, {\color{black}we check that convergence is achieved by estimating $\alpha$ and $\beta$, as} introduced in Section \ref{subsec:mlmc_theory}.
\subsubsection{Pilot Run -- Reduced Basis and Sample Size Determination} \label{subsubsec:pilot}

As in the case of MLMC, the goal of the pilot run is to find estimates of $\mathbb{E}[Y_\ell]$  and $\mathbb{V}[Y_\ell]$, to ensure level $L$ is fine enough such that the discretization error is bounded by $\varepsilon^2/2$, and to determine the sample sizes $\tilde{N}_\ell$ and $N'_{\ell}$. However, determining $\tilde{N}_\ell$ and $N'_{\ell}$ requires the estimation of $\rho^2_{\ell}$ and the derivation of the reduced basis $\bm{U}^c_{\ell}$ (along with the associated interpolation rules) to construct the control variates $Z_\ell$. These are specific to this MLCV approach and will be additionally performed during the pilot run stage summarized in Algorithm \ref{alg:pilot}.

In particular, a small set of $N_p\ge m_{L-1}$ samples of $\bm \xi$ is generated and applied to level $\ell-1$ and $\ell$ models to obtain samples of $Y_\ell$, as well as estimates of $\mathbb{E}[Y_\ell]$  and $\mathbb{V}[Y_\ell]$. We note that $N_p$ may vary from one level to the other, provided that some prior knowledge of $\mathbb{V}[Y_\ell]$ is available. As discussed in Section \ref{subsec:z_mid}, a reduced basis $\bm{U}^c_{\ell-1}$ and an associated interpolation rule for the level $\ell-1$ solution vector $\q_{\ell-1}$ is found via the ID of the coarse grid data matrix $\bm{U}_{\ell-1}$. The level $\ell$ reduced basis $\bm{U}^c_{\ell}$ is generated by simulating the level $\ell$ model at the samples of $\bm{\xi}$ identified by the level $\ell-1$ ID and used -- in conjunction with the same interpolation rule -- to acquire samples of $\q_{\ell}^{ID}$, and subsequently $Q^{ID}_\ell$ and $Z_\ell$. These bases are stored and reused in the MLCV runs of Section \ref{sec:mlcv_runs}. Following this, $\rho^{2}_\ell$ and $\tilde{N}_{\ell}/N'_{\ell}$ are computed from (\ref{r2}) and (\ref{npstar3}), respectively. After each level the change in the sample means of $Y_\ell$ must be determined to verify that the intended discretization accuracy is met. If not, this process is repeated with larger values of $L$. If the value of $L$ is sufficient, the pilot run ends and the values of $\tilde{N}_\ell$ and $N'_{\ell}$ are determined from (\ref{nlcv}) and (\ref{r2}), respectively.
{\color{black}
\begin{remark}
While the variance and covariance are estimated from the pilot run, these values, in practice, may be updated during the main MLCV run described next.
\end{remark}}

\RestyleAlgo{boxruled}
\SetKwInOut{Input}{input}
\SetKwInOut{Output}{output}
\SetKw{Initialize}{initialize}
\SetKw{Return}{return}
\SetKw{Break}{break}

\begin{algorithm}
\For{$\ell=0,\dots,L$}{
\uIf{$\ell=0$}{
$\{ {\bm \xi}^{(i)} \}_{i=1}^{N_p} \longrightarrow \{ Q_{\ell}^{(i)} \}_{i=1}^{N_p} \longrightarrow \{ Y_{\ell}^{(i)} \}_{i=1}^{N_p}$} 
\Else{
$\{ {\bm \xi}^{(i)} \}_{i=1}^{N_p}\longrightarrow \{\q_{\ell-1}^{(i)}\}_{i=1}^{N_p},  \{\q_{\ell}^{(i)}\}_{i=1}^{N_p}\longrightarrow \{ Q_{\ell-1}^{(i)} \}_{i=1}^{N_p},\{ Q_{\ell}^{(i)} \}_{i=1}^{N_p} \longrightarrow \{ Y_{\ell}^{(i)} \}_{i=1}^{N_p}\ $ \vspace{2mm}\\ 

$\bm{U}_{\ell-1} = [\q^{(1)}_{\ell-1}\ \cdots \ \q^{(N_p)}_{\ell-1}]\xrightarrow[]{\text{ID via } (\ref{eq:CGmid})}  \bm{U}_{\ell-1}^c\, \bm{C}_{\ell-1} , \{{\bm \xi}^{(i_k)} \}_{i=1}^r$\vspace{2mm}\\

$ \{{\bm \xi}^{(i_k)} \}_{i=1}^r \longrightarrow \bm{U}_\ell^c\ ;\ \bm{U}^{ID}_{\ell} = \bm{U}_{\ell}^c\bm{C}_{\ell-1}  \longrightarrow \{ Q^{ID (i)}_{\ell} \}_{i=1}^{N_p} \longrightarrow  \{ Z_{\ell}^{(i)} \}_{i=1}^{N_p} $\vspace{2mm}\\

Calculate $ \rho^2_\ell$ and $\tilde{N}_{\ell}/N'_{\ell}$, respectively, via (\ref{r2}) and (\ref{npstar3})\vspace{2mm}\\
}
Estimate $\mathbb{V}[Y_{\ell}]$\\

}
\For{$\ell=0,\dots,L$}{
Calculate $\tilde{N}_\ell$ using (\ref{nlcv})\\
Calculate $N'_\ell$ using (\ref{npstar3})}
\protect\caption{MLCV Reduced Basis Identification and Sample Size Determination\label{alg:pilot}}
\end{algorithm}

\subsubsection{Main MLCV Run}
\label{sec:mlcv_runs}

After the total number of samples $\tilde{N}_\ell$ and $N'_\ell$ for each level are estimated, the full MLCV run is completed following Algorithm \ref{alg:mlcv}. Much like the pilot run, $\tilde{N}_\ell$ realizations of $Y^{(i)}_\ell$ are collected from the level $\ell$ and $\ell-1$ models. Using the $\tilde{N}_\ell$ samples of $\q_{\ell-1}$, and the level $\ell-1$ reduced basis $\bm{U}^c_{\ell-1}$ from the pilot run, least squares (\ref{eqn:least_squares}) is performed on each sample to find the coefficient vectors $\c_{\ell-1}(\bm\xi)$. Given this interpolation rule and the corresponding level $\ell$ basis $\bm{U}^c_{\ell}$, the samples of $\q_{\ell}^{ID}$ (hence $Q_\ell^{ID}$) and subsequently $Z_\ell^{(i)}$ are generated using (\ref{eq:FGmid}) and (\ref{eq:Z}), respectively. Following these, $\bar{Z}_\ell$ is computed from (\ref{eq:z_bar}) using $N'_\ell$ independent samples of $Z_\ell$ drawn as described above. After calculating $\theta^{*}_\ell$ in (\ref{beta}), we have all the components of the MLCV estimator, and use (\ref{cvem}) to find $\hat{W}_\ell$ and lastly $\hat{Q}_{L}^{MLCV}$.

\begin{algorithm}
\For{$\ell=0,\dots,L$}{
\uIf{$\ell=0$}{
$\{ {\bm \xi}^{(i)} \}_{i=1}^{\tilde{N}_\ell}\longrightarrow \{ Q_{\ell}^{(i)} \}_{i=1}^{\tilde{N}_\ell} \longrightarrow \{ Y_{\ell}^{(i)} \}_{i=1}^{\tilde{N}_\ell}\longrightarrow \hat{W}_\ell = \hat{Y}_\ell$}
\Else{

\vspace{3mm}

$\%\ \bar{Z}_{\ell}$ calculation:\\ 

$\{ {\bm \xi}^{(i)} \}_{i=1}^{N'_{\ell}}\longrightarrow \{\q_{\ell-1}^{(i)}\}_{i=1}^{N'_\ell}\longrightarrow \{ Q_{\ell-1}^{(i)} \}_{i=1}^{N'_\ell}$\vspace{2mm}\\

$\c_{\ell-1}^{(i)}= \min\limits_{\hat{\mathbf{c}}\in \mathbb{R}^r} \Vert \bm{U}^c_{\ell-1}\ \hat{\mathbf{c}} -\q_{\ell-1}^{(i)} \Vert\ \text{for }i=1,\dots,N'_\ell$\vspace{2mm}\\

$\q^{ID (i)}_{\ell} = \bm{U}_{\ell}^c\c_{\ell-1}^{(i)}\ \text{for } i=1,\dots,N'_\ell  \longrightarrow \{ Q^{ID (i)}_{\ell} \}_{i=1}^{N'_{\ell}} \longrightarrow  \{ Z_{\ell}^{(i)} \}_{i=1}^{N'_{\ell}}\longrightarrow \bar{Z}_\ell$\vspace{4mm}\\

$\%\ Z_{\ell}$ sampling:\\ 

$\{ {\bm \xi}^{(i)} \}_{i=1}^{\tilde{N}_\ell}\ (\text{independent of }\{ {\bm \xi}^{(i)} \}_{i=1}^{N'_{\ell}}\ \text{above})\longrightarrow \{\q_{\ell-1}^{(i)}\}_{i=1}^{\tilde{N}_\ell},  \{\q_{\ell}^{(i)}\}_{i=1}^{\tilde{N}_\ell}\longrightarrow \{ Q_{\ell-1}^{(i)} \}_{i=1}^{\tilde{N}_\ell},\{ Q_{\ell}^{(i)} \}_{i=1}^{\tilde{N}_\ell} \longrightarrow \{ Y_{\ell}^{(i)} \}_{i=1}^{\tilde{N}_\ell}\ $ \vspace{2mm}\\ 

$\c_{\ell-1}^{(i)}= \min\limits_{\hat{\mathbf{c}}\in \mathbb{R}^r} \Vert \bm{U}^c_{\ell-1}\ \hat{\mathbf{c}} -\q_{\ell-1}^{(i)} \Vert\ \text{for } i=1,\dots,\tilde{N}_\ell$\vspace{2mm}\\

$\q^{ID (i)}_{\ell} = \bm{U}_{\ell}^c\c_{\ell-1}^{(i)}\ \text{for } i=1,\dots,\tilde{N}_\ell  \longrightarrow \{ Q^{ID (i)}_{\ell} \}_{i=1}^{\tilde{N}_{\ell}} \longrightarrow  \{ Z_{\ell}^{(i)} \}_{i=1}^{\tilde{N}_{\ell}}\ $\vspace{4mm}\\

$\%\ W_{\ell}$ sampling:\\ 
 $\{W^{(i)}_{\ell}\}_{i=1}^{\tilde{N}_\ell}$ from (\ref{cvem}) $\longrightarrow \hat{W}_\ell$ \vspace{2mm}\\

}
}
Set $\hat{Q}^{MLCV}_L = \sum_{\ell=0}^L  \hat{W}_{\ell}$

\protect\caption{Main MLCV Run\label{alg:mlcv}}
\end{algorithm}

\subsection{MLCV Cost Breakdown}
\label{subsec:mlcv_cost}

After determining the total number of samples required to optimize the cost with the MSE bound as a constraint, the total cost of implementing MLMC is given in (\ref{eq:mlmc_totalcost}). By modifying MLMC through the use of the control variate $Z_\ell$, we are able to achieve further reduction in the MSE of the estimator $\hat{W}_\ell$, as long as $Z_\ell$ is sufficiently correlated with $Y_\ell$. In doing so, the sample size for $Y_\ell$ and thus $Z_\ell$ needed to meet the sampling error bound, while optimizing the total cost, is reduced. However, sampling $Z_\ell$ and estimating its expectation, as required in (\ref{eqn:control_eq_mlcv}), lead to additional computational cost that we quantify next. 

\paragraph{\it Cost of pilot run} Recall that the construction of $Z_\ell$ is based on the identification of the reduced basis $\bm{U}^c_{\ell-1}$ (using which $\bm{U}^c_{\ell}$ is generated) and the associated least squares coefficients $\c_{\ell-1}$ computed from (\ref{eqn:least_squares}). Following Algorithm \ref{alg:pilot}, the former is based on the ID of the coarse data matrix $\bm{U}_{\ell-1}$ consisting of $N_p$ pilot samples of $\q_{\ell-1}\in\mathbb{R}^{m_{\ell-1}}$. From \cite{Martinsson11}, the cost to implement a rank $r$ ID on $\bm{U}_{\ell-1}$ is $\mathcal{O}(m_{\ell-1}\tilde{N}_{\ell} n_r)$, where $n_r<\tilde{N}_{\ell}$. This means that the cost per sample of $Z_\ell$ is $\mathcal{O}(m_{\ell-1} n_r)$. In many scenarios of practical interest, the QoI $Q$ depends on a $\q$ consisting of a considerably smaller set of solution degrees of freedom, e.g., average outflow temperature or lift/drag on an airfoil, as opposed to those of the entire solution. In such cases, $m_{\ell-1}\ll M_{\ell-1}$ and we may ignore the cost of performing the ID of $\bm{U}_{\ell-1}$. {\color{black}Assuming that (i) the factorization to solve (\ref{eqn:least_squares}) is formed only once, e.g, using a reduced QR decomposition with cost $\mathcal{O}(r^2m_{\ell-1})$,} (ii) $r\ll m_{\ell}$, and (iii) $\tilde{N}_{\ell}\gg 1$, the cost of solving for the coefficients $\c_{\ell-1}$ for each sample {\color{black}and applying them to $\bm{U}^c_{\ell}$ to obtain $\q_{\ell}^{ID}$ is $\mathcal{O}(r m_{\ell})$}, which again may be ignored when $m_{\ell}\ll M_{\ell}$. From the $N_p$ pilot samples, $r$ of them form the reduced bases $\bm{U}^c_{\ell-1}$ and $\bm{U}^c_{\ell}$ the rest will be recycled in the main MLCV run, assuming $N_p\le \tilde{N}_{\ell}$. Therefore, the net cost of performing the pilot run on each level is the same as the cost of generating $r$ samples of level $\ell$ and $\ell-1$ solutions to form the reduced bases $\bm{U}^c_{\ell-1}$ and $\bm{U}^c_{\ell}$. This gives rise to the total cost of $\sum_{l=1}^{L}r\left(\mathcal{C}(Q_{\ell-1})+\mathcal{C}(Q_{\ell})\right)$ for the pilot run.

\paragraph{\it Cost of main MLCV run} Sampling $Z_\ell$ in (\ref{eqn:control_mlcv_zbar}) entails using the $\tilde{N}_\ell$ samples of $\q_{\ell-1}$ drawn for $Y_\ell$ and performing least squares regression to generate the approximation (\ref{eqn:rank_r_q}). As discussed previously, these least squares solves have negligible cost when $m_\ell\ll M_\ell$. Therefore, there is no additional cost for sampling $Z_\ell$ in (\ref{eqn:control_mlcv_zbar}) relative to its MLMC counterpart. With a similar argument, sampling $Z_\ell$ in (\ref{eqn:control_mlcv_zbar}) to compute $\bar{Z}_\ell$ requires only generating $N'_\ell$ independent samples of $\q_{\ell-1}$. Therefore, the cost of the main run is $\tilde{N}_0\mathcal{C}(Q_0)+\sum_{\ell=1}^L  \tilde{N}_{\ell}\left(\mathcal{C}(Q_{\ell-1})+\mathcal{C}(Q_{\ell})\right) + \sum_{\ell =1}^L N'_\ell\mathcal{C}(Q_{\ell-1})$.   

\paragraph{\it Total cost} Adding the cost of pilot and main MLCV runs, we arrive at the total cost of MLCV given by
\begin{equation}
\label{eq:cost_mlcv}
\mathcal{C}(\hat{Q}^{MLCV}_L) = \tilde{N}_0\mathcal{C}(Q_0)+\sum\limits_{\ell=1}^L  (\tilde{N}_{\ell}+r)\left(\mathcal{C}(Q_{\ell-1})+\mathcal{C}(Q_{\ell})\right) + \sum_{\ell =1}^L N'_\ell\mathcal{C}(Q_{\ell-1}).
\end{equation}
\subsection{Discussion on Use of MLCV}
\label{sec:discussion}

Before proceeding to the numerical experiments and results, we first discuss the types of problems that benefit from the application of MLCV. When deciding to apply a MC based method to solve a high-dimensional problem in uncertainty quantification, the practitioner must consider the QoI's variance -- more specifically coefficient of variation (COV) -- as well as the cost of simulating the model on a fine enough discretization. For problems with small COV and cheap-to-evaluate models, it can be the case that standard MC will perform as well, or may outperform MLMC. On the other hand, the practitioner will likely decide to use MLMC over MC if the QoI shows a large COV and the solver exhibits a fast cost growth between refined grids ($\mathcal{C}_\ell \gg M_\ell$). Since MLCV is a multilevel method, it benefits from the same traits as MLMC: a large COV and a fast cost growth. However, unlike MLMC, MLCV relies on a low-rank representation of the solution, in the sense defined in Section \ref{subsec:z_mid}, thus making the approach most effective in outperforming MLMC on problems exhibiting small solution ranks on all the levels. When this occurs we find that $Q^{ID}_\ell$ approximates the fine grid data $Q_\ell$ accurately, and that the value for $\rho^2_\ell$ is much closer to one, thus resulting in an improved MSE for $\hat{W}_\ell$.

A final comment must be made regarding the total cost of MLCV. In the case that significant MSE reduction for $\hat{W}_\ell$ does occur, we are not necessarily guaranteed cost reduction. In (\ref{eq:cost_mlcv}) the number of samples needed on levels $\ell>0$ is always larger than $r$. In MLMC, this constraint is not present (see (\ref{eq:mlmc_totalcost})). For scenarios where the data exhibits a small variance or small values of $\varepsilon$ are provided as the desired tolerance, the resulting number of samples on the finer levels may be comparable to $r$. In such cases, MLCV may not outperform MLMC. This suggests MLCV is more beneficial for smaller selections of $\varepsilon$.

\section{Numerical Results}
\label{sec:results}

In this section we consider two test cases to compare the resulting error and cost estimates for MLMC and MLCV. In the first case we consider the generalized eigenvalue problem for finding the frequency of the first natural mode of vibration of an L-shaped elastic structure. For the second case we consider a thermally driven cavity flow with stochastic boundary temperatures previously studied in \cite{LeMaitre10, Rubio02,Peng14,Hampton15a}. 

Following each case, we present several results when comparing MLCV and MLMC for both tests. In particular, we consider the convergence of $\rho_\ell^2$ in (\ref{r2}), and subsequently the MSE at each level, which controls the sampling error of MLCV. For values of $\rho_\ell^2$ close to one, a significant reduction in the MSE of $\hat{W}_\ell$ can be observed, and thus a reduction in the number $\tilde{N}_\ell$ of required samples per level. We further provide results on the total cost and relative error of MLCV and MLMC. The total cost, as a function of $\varepsilon$, is determined from (\ref{nlcv}), (\ref{npstar3}), and (\ref{eq:cost_mlcv}). For the eigenvalue problem, the cost is defined to be $\mathcal{C}_\ell = M_\ell$, while the thermally driven flow problem cost is approximated by an average CPU time. The final result is the relative error in estimating the mean of QoI -- to be specified -- on all discretization levels. The errors are generated based on a reference solution $\hat{Q}^{\text{ref}}_L$ obtained via sparse polynomial chaos expansion \cite{Doostan11a, Hampton15a} on the finest available discretization for each model. Specifically, 
\begin{equation}\label{eq:rel_err_mlmc}
e^{MLMC}_\ell=\frac{ \left|\sum\limits_{k=0}^{\ell} \hat{Y}_k  - \hat{Q}^{\text{ref}}_L\right|}{\left| \hat{Q}^{\text{ref}}_L\right|}
\end{equation}
and
\begin{equation}\label{eq:rel_err_mlcv}
e^{MLCV}_\ell= \frac{\left|\sum\limits_{k=0}^{\ell} \hat{W}_k  - \hat{Q}^{\text{ref}}_L\right|}{\left| \hat{Q}^{\text{ref}}_L\right|},
\end{equation}
for $\ell=1,\dots, L$.
\subsection{Test 1: Natural Frequency of a 2D Linear Elasticity Problem}

\begin{figure}
\centering
\subfloat[Schematic of domain]{\includegraphics[trim= 48mm 35mm 98mm 30mm , height=1.2in]{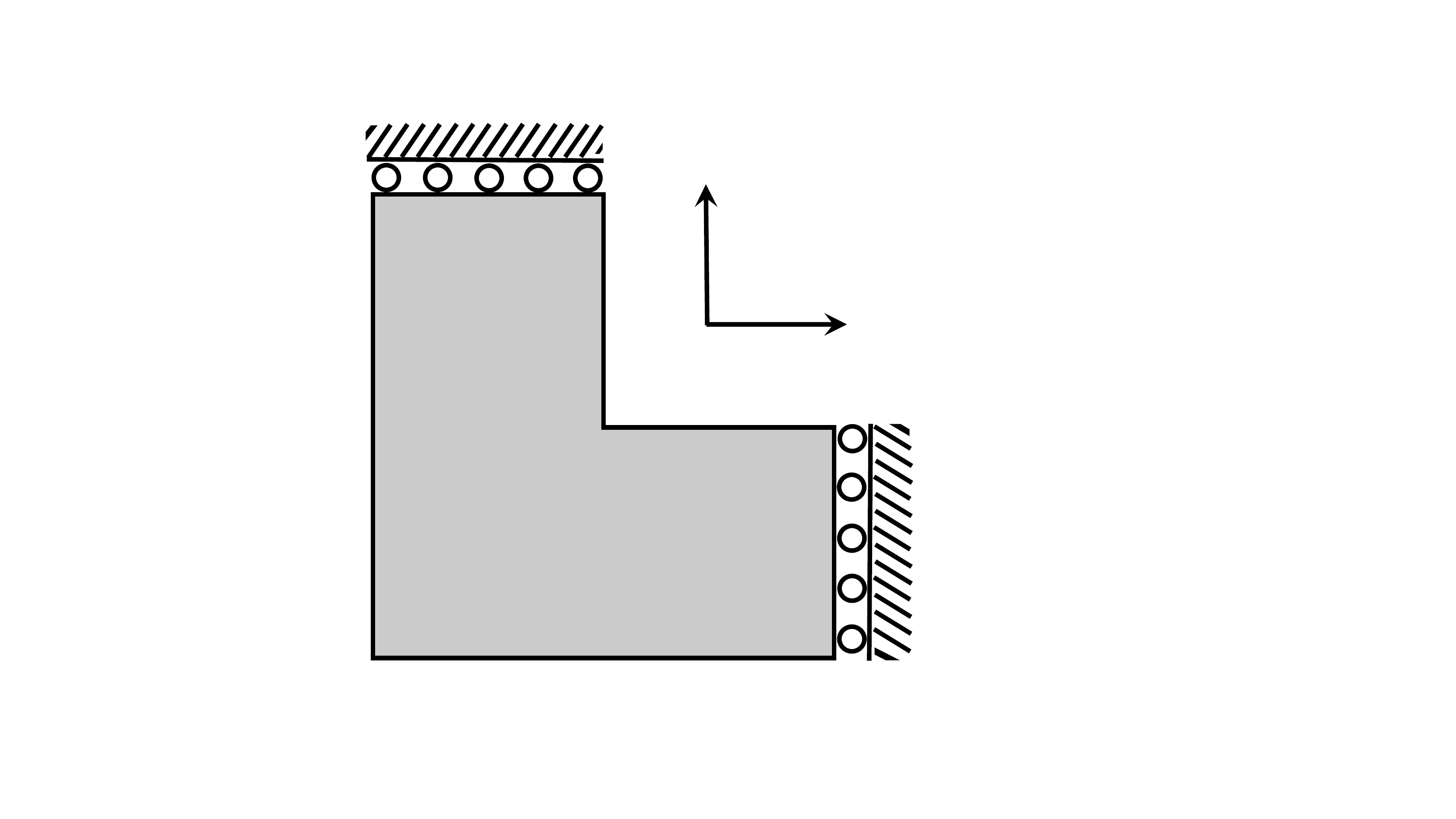}
\put(-28,50){\scriptsize $x$}
\put(-60,74){\scriptsize  $y$}
\put(-67,43){\tiny  $(0,0)$}
\put(-115,-5){\tiny  $(-1,-1)$}
}
\subfloat[$\ell =0$]{\includegraphics[trim= 25mm 20mm 25mm 15mm,height=1.2in]{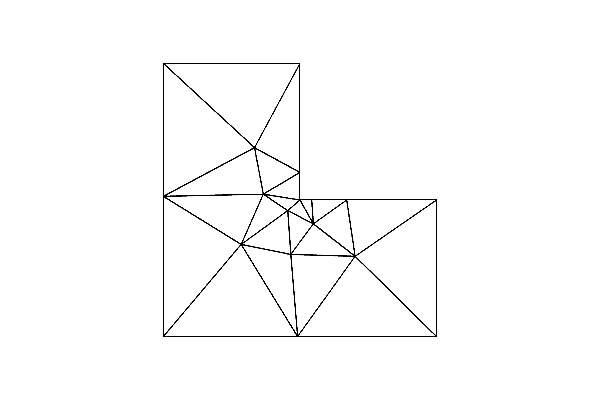}}
\subfloat[$\ell =1$]{\includegraphics[trim= 25mm 20mm 25mm 15mm,height=1.2in]{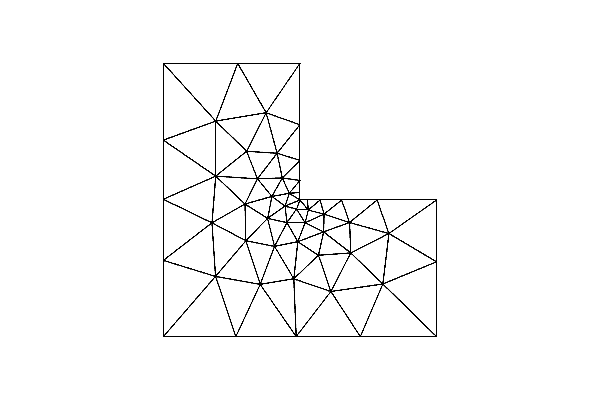}}
\hspace{\textwidth}
\subfloat[$\ell =2$]{\includegraphics[trim= 25mm 20mm 25mm 15mm,height=1.2in]{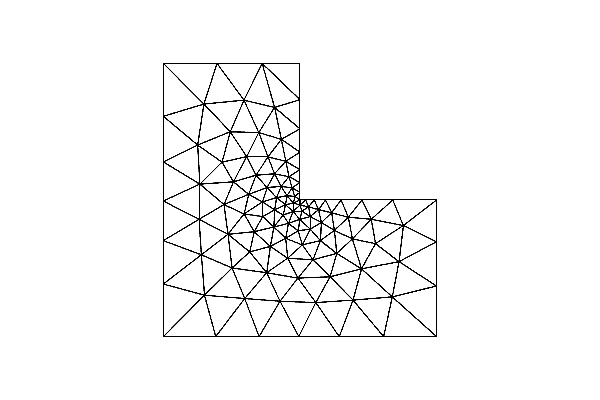}}
\subfloat[$\ell =3$]{\includegraphics[trim= 25mm 20mm 25mm 15mm,height=1.2in]{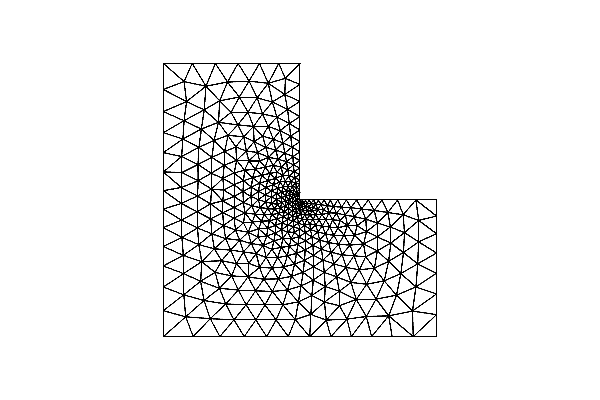}}
\subfloat[$\ell =4$]{\includegraphics[trim= 25mm 20mm 25mm 15mm,height=1.2in]{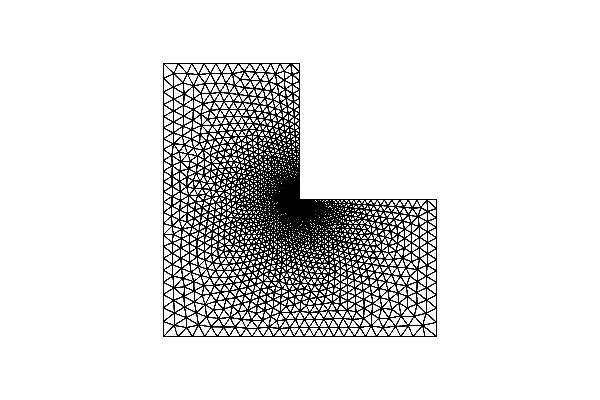}}
\caption{(a) Schematic of the L-shaped domain. (b)-(f) Meshes on the five different levels with increasing resolution.\label{fig:Lmesh}}
\end{figure}

For the first test, we seek to find the expected value of the smallest vibration frequency, $\omega $, of the linear elasticity problem on an L-shape domain $\mathcal{D}$ as shown in Figure \ref{fig:Lmesh} (a). We consider zero Dirichlet boundary conditions on the top and right sides of the domain along $y$ and $x$ directions, respectively. {\color{black}The Young's modulus of the medium, given by the log-normal random field
\[
E(\bm x, \bm \xi) = \bar{E} + \text{exp}\left( G(\bm x, \bm \xi)   \right),\qquad \bm x\in\mathcal{D},
\]
is the source of uncertainty in this test. Here, $G(\bm x, \bm \xi)$ is a Gaussian random field represented by the Karhunen-Lo\`{e}ve expansion
\[
G(\bm x, \bm \xi) = \sum\limits_{i=1}^{d} \sqrt{\lambda _i} \phi _i (\bm x) \xi _i,
\]
where the random variables} $\xi _i$ are independent standard Gaussian and $\lambda_i$ are the $d$ largest eigenvalues with corresponding eigenfunctions $\phi _i(\bm x)$ determined by the Gaussian covariance function
\begin{equation}\label{eq:cov1}
\mathcal{K}(\bm x _1,\bm x _2) = \sigma ^2\text{exp}\left(- \frac{\Vert \bm x _1 - \bm x _2  \Vert_2^2}{l}\right).\nonumber
\end{equation}
In our experiments, we set $\bar{E}= 0.1$, $d=36$, $\sigma^2 = 0.33$, and $l = 0.33$, and assume the medium has a unit density.

To compute $\omega$, we consider the solution of the random eigenvalue problem 
\begin{equation}
\label{eq:eigen_problem}
\bm K(\bm\xi) \q(\bm\xi) = \omega ^2(\bm\xi) \bm M \q(\bm\xi),
\end{equation}
where $\bm K(\bm\xi)$ and $\bm M$ are, respectively, the stiffness and mass matrices associated with finite element discretization of the linear elasticity equations on uniform triangular meshes of varying size. Additionally, $\q(\bm\xi)$ is the natural vibration mode (eigenvector) corresponding to the eigenvalue $\omega^2(\bm\xi)$ given by
\begin{equation}
\label{eq:eval}
\omega^2(\bm\xi) = \frac{\q(\bm\xi)^T\bm K(\bm\xi) \q(\bm\xi)}{ \q(\bm\xi)^T\bm M \q(\bm\xi)}.
\end{equation}
Here, QoI is the smallest frequency $\omega$ satisfying (\ref{eq:eigen_problem}) and is a functional of $\q(\bm\xi)$, i.e. $Q = Q\left(\q(\bm\xi)\right)$, through (\ref{eq:eval}). With the given uncertainty, the COV of $Q$ is approximately $15\%$.

The finite element discretization is done using FENiCS \cite{LoggMardalEtAl2012a} with five different meshes of varying size, as displayed in Figure \ref{fig:Lmesh} (b)-(f). The number of degrees of freedom of the meshes are $M_0 = 36$, $M_1 = 106$, $M_2 = 212$, $M_3 = 856$, and $M_4 = 3296$. The resulting eigenvector $\q_\ell$, from each level $\ell$ simulation, has $M_\ell$ degrees of freedom. After eigenvectors $\q_\ell^{ID}$ are found, the corresponding (saved) stiffness and mass matrices, $\bm K$ and $\bm M$, are used in (\ref{eq:eval}) to generate $Q_\ell^{ID}$ and the control variates $Z_\ell$. 

For the application of MLMC and MLCV we consider two cases to test the relative performance of these methods when altering the difference in the number of degrees of freedom between adjacent levels. In the first case, {\it A}, we apply MLMC and MLCV to all five levels. In the second case, {\it B}, we consider only using three of the five meshes from levels $\ell = 0,2,4$. This is worth testing, as both methods rely on the differences between QoIs on adjacent levels. The results between {\it A} and {\it B} will indicate which method can adjust to the larger gap in the number of degrees of freedom between levels for this problem. For the rank selected on each level, i.e., the size of the reduced basis, we use $r =10$ for all levels, with the exception of $r = 15$ on the finest level in case {\it A}, where the increase was required to have $\rho_L^2 > 0.90$. 

\subsubsection{Results of Test 1}

\begin{center}
\begin{figure}
\resizebox{.5\textwidth}{!}{\includegraphics[trim= 10mm 55mm 10mm 60mm]{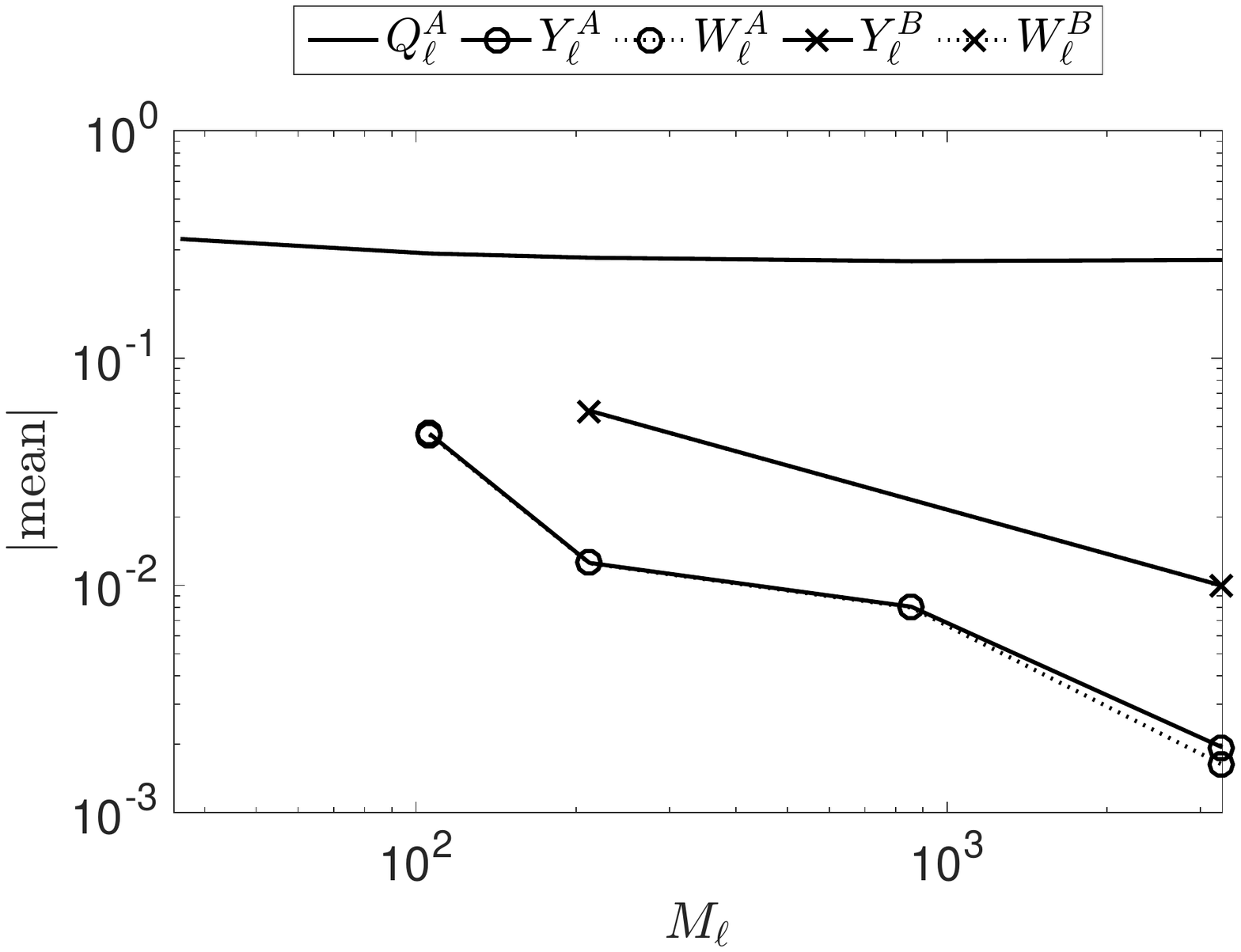}}
\resizebox{.5\textwidth}{!}{\includegraphics[trim= 10mm 55mm 10mm 60mm]{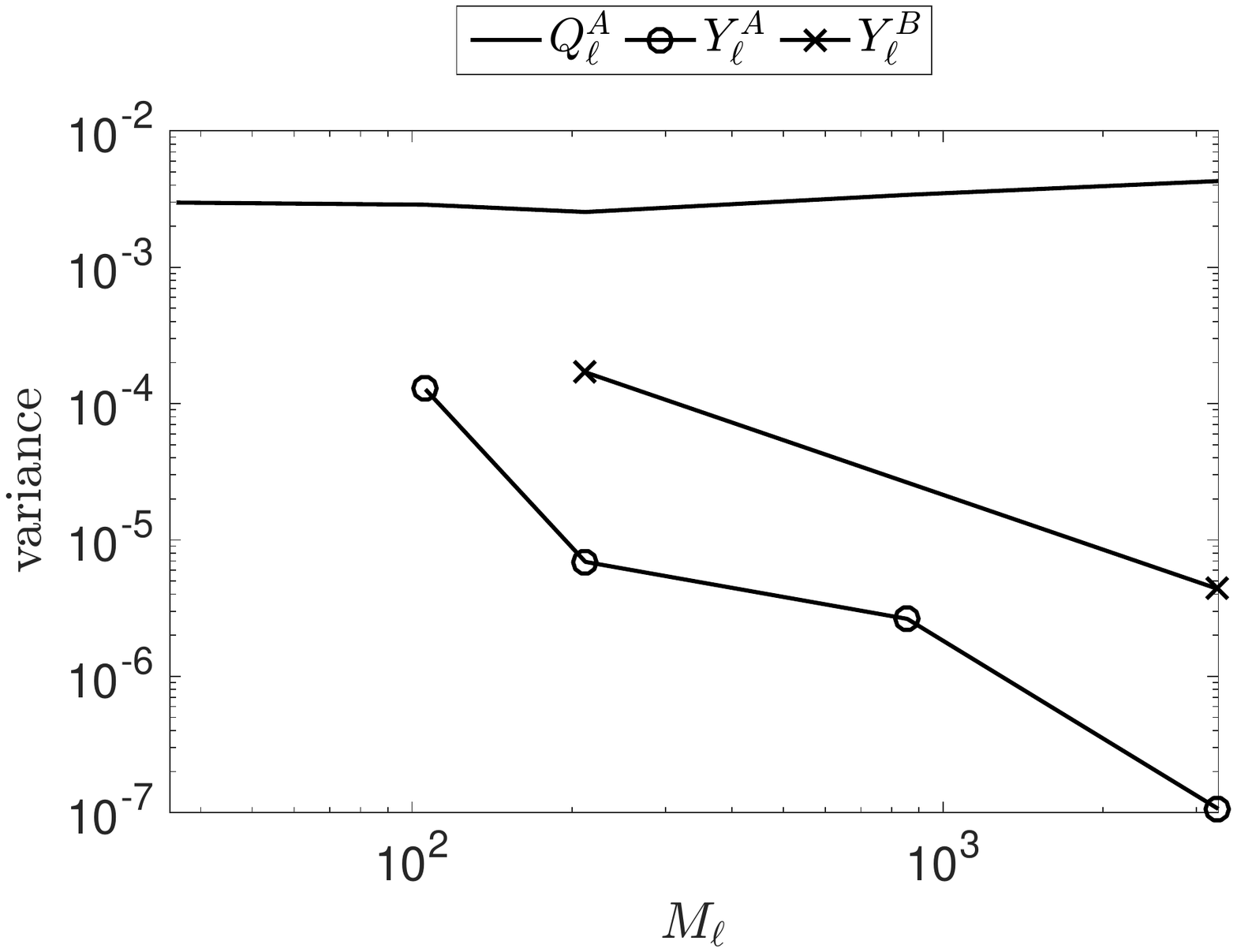}}
\captionof{figure}{(Left) Convergence of the sample mean of $Y_\ell$ and $W_\ell$, while that of $Q_\ell$ remains relatively constant. (Right) Convergence of the sample variance of $Y_\ell$ in comparison to that of $Q_\ell$. Superscripts $A$ and $B$ refer to the case $A$ and $B$ of the L-shaped domain problem, respectively.\label{fig:EVL}}
\end{figure}
\end{center}

For the results presented in this section, the ``o" markers denote case {\it A} data and the ``$\times$" markers denote case {\it B} data. The first set of results are related to the MLMC convergence guarantee in Section \ref{subsec:mlmc_theory}. Figure \ref{fig:EVL} (left) shows the sample mean of $Q_\ell$, $Y_\ell$ and $W_\ell$ as a function of the number of degrees of freedom on each level, $M_\ell$. We note that the sample mean of $Q_\ell$ remains relatively constant, while those of $Y_\ell$ and $W_\ell$ decay at the same rate. The data displayed allows us to estimate the values of $\alpha$. Based on the slope, we have $\vert\mathbb{E}[Q_\ell-Q_{\ell-1}]\vert\approx M_\ell^{-0.92} $ and $\vert\mathbb{E}[Q_\ell-Q_{\ell-1}]\vert\approx M_\ell^{-0.64} $ for case {\it A} and {\it B}, respectively. Thus it follows that $\alpha _A \approx 0.92$ and $\alpha _B \approx 0.64$. Figure \ref{fig:EVL} (right) shows the sample variance of $Q_\ell$ and $Y_\ell$ as a function of the degrees of freedom on each level, $M_\ell$. The data displayed allows us to estimate the values of $\beta$. From Figure \ref{fig:EVL} (right) we have $\mathbb{V}[Q_\ell - Q_{\ell-1}]\approx M_\ell^{-1.7}$ and $\mathbb{V}[Q_\ell - Q_{\ell-1}]\approx M_\ell^{-1.3}$, for case {\it A} and {\it B}, respectively. This implies that $\beta _A \approx 1.7$ and $\beta _B \approx 1.3$. The cost of the FE solver, as well as the cost to compute $\omega _1^2$, indicate that $\gamma \approx 1$, i.e. $\mathcal{C}_{\ell} \lesssim M_{\ell}$, suggesting that for case {\it A} and {\it B} we may consider a tolerance of $\varepsilon \sim \mathcal{O}(0.0005)$. We remark that, while the value of $\alpha_B$ indicates the discretization error cannot meet the specified tolerance, we argue that this is an estimation of $\mathbb{E}[Q_L-Q]$, and that this approximation of $\alpha$ is better obtained from test {\it A} results.

\begin{center}
\begin{figure}[ht]
\resizebox{.5\textwidth}{!}{\includegraphics[trim= 10mm 55mm 10mm 60mm]{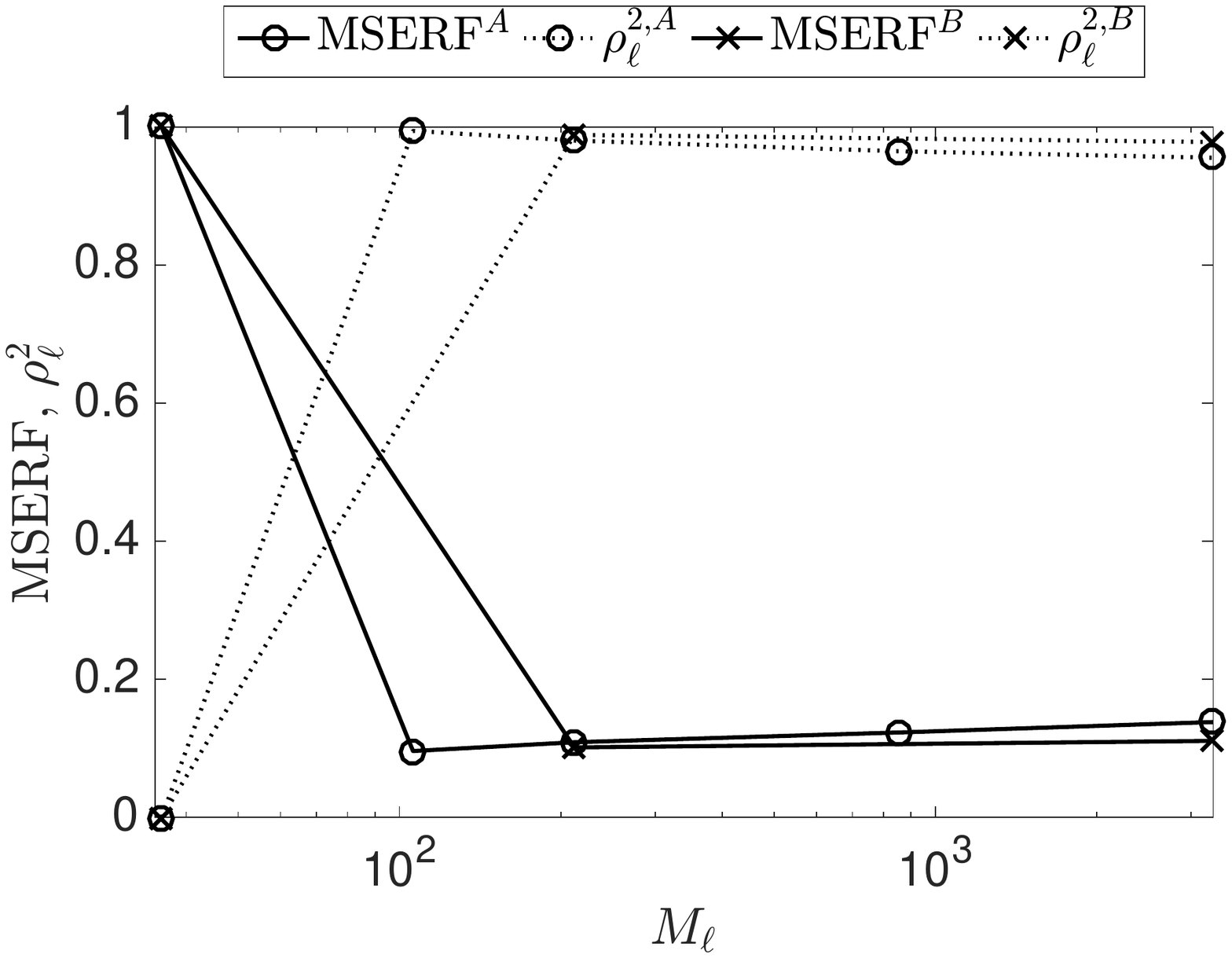}}
\resizebox{.5\textwidth}{!}{\includegraphics[trim= 10mm 55mm 10mm 60mm]{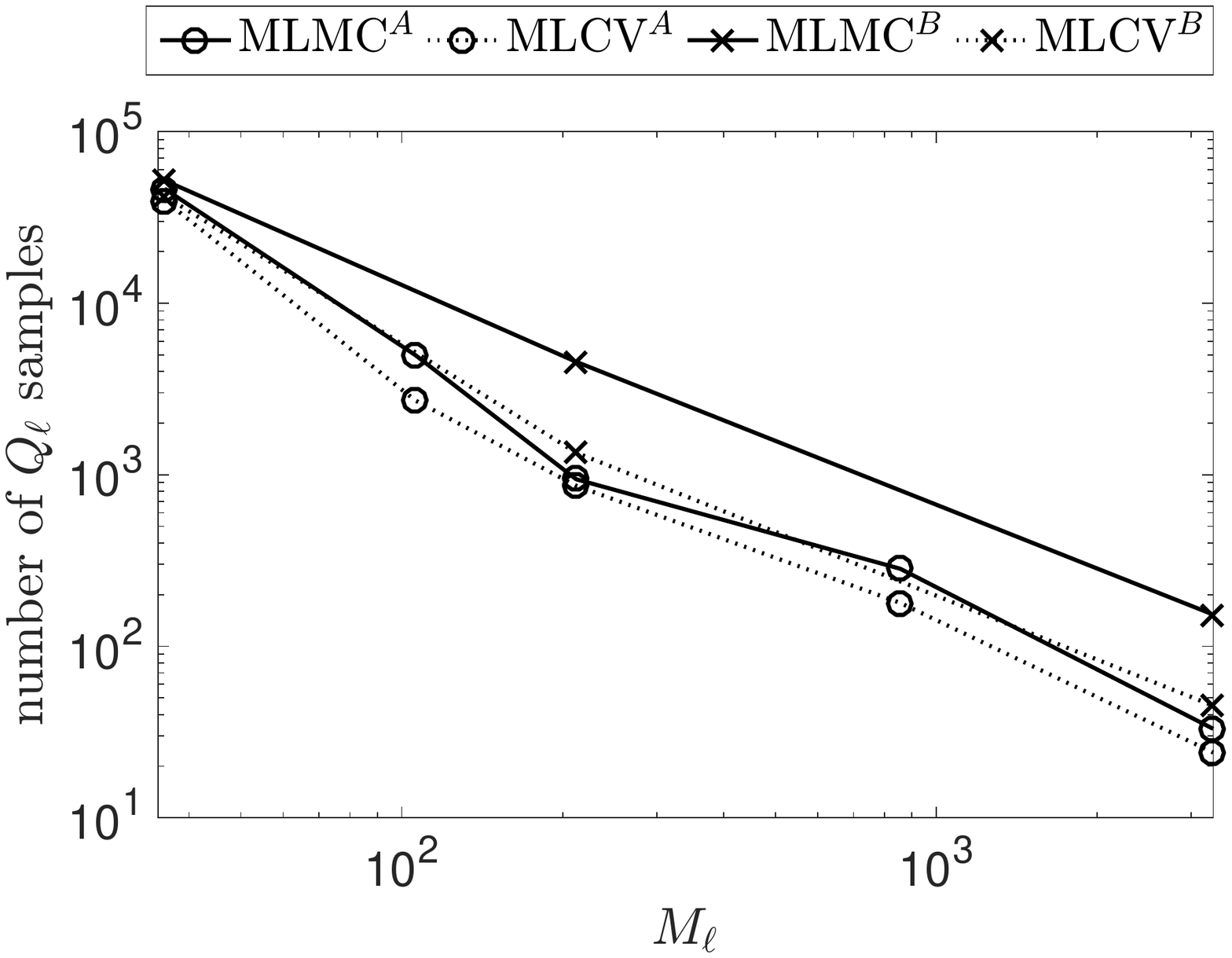}}
\captionof{figure}{(Left) MSE reduction factor (MSERF) between $\hat{W}_\ell$ and $\hat{Y}_\ell$, as well as the value of $\rho_\ell^2$ on each level. (Right) Total number of samples needed to achieve the sampling error $\varepsilon=0.0005$ for MLMC and MLCV. MLCV sample size includes the number required to determine $\bar{Z}_{\ell}$.  Superscripts $A$ and $B$ refer to the case $A$ and $B$ of the L-shaped domain problem, respectively. \label{fig:MSEL}}
\end{figure}
\end{center}

Next we consider the MSE reduction in Figure \ref{fig:MSEL}. Figure \ref{fig:MSEL} (left) shows the increasing values of $\rho_\ell^2$ as a function of $\ell$, as well as the decreasing values of the MSE reduction factor. For both cases {\it A} and {\it B}, $\rho_\ell^2>0.95$ for $\ell >0$. This is ideal, as this reduction in the MSE means that fewer samples are required to obtain the MSE tolerance. Figure \ref{fig:MSEL} (right) displays the number of samples required on each level for both methods when using $\varepsilon = 0.0005$. Note that the number of samples for MLCV includes those used to calculate $\bar{Z}_\ell$. For case {\it A} the number of samples required for MLCV is slightly smaller than that of MLMC. A more significant reduction in the number of samples needed for MLCV is observed for case {\it B}. This suggests that MLCV is more adaptable when large gaps are added between $M_\ell$ on adjacent levels.


To compare the costs for these two methods, we turn to Table \ref{t:1}, which displays the cost estimates for several cases as a function of prescribed $\varepsilon$ based on (\ref{eq:mlmc_totalcost}) and (\ref{eq:cost_mlcv}), where $\mathcal{C}_\ell=M_\ell$. For comparison, the cost of MC on the finest level is also reported. For small values of $\varepsilon$, an improvement in cost is observed for MLCV. When requiring a tolerance of $\varepsilon = 0.0005$ we observe a cost ratio of $75\%$ for case {\it A} and $58\%$ for case {\it B}. Comparing the total costs of these two cases, we see that the MLCV method for case {\it B} outperforms case {\it A}. For MLMC, it is the opposite. By adding larger gaps between levels, the performance of MLMC decays, while that of MLCV improves. When increasing the tolerance, there is a reduction in the cost gain of MLCV, as fewer samples are needed to maintain the MSE tolerance. The cost of the $r$ samples for the basis is diminishing the success of the MLCV method. When $\varepsilon > 0.003$, the cost of performing MLCV is greater than the cost of performing MLMC. 

\begin{table}[h]
\centering
\caption{Cost of MC, MLMC, and MLCV for Test 1 problem. The difference in cost of MC for $\varepsilon = 0.0005$ is due to the small number of samples used.\label{t:1}}
\begin{tabular}{lllllllllll}
$\varepsilon$ & Levels & Cost MC &Cost MLMC&Cost MLCV & $\frac{\text{Cost MLCV} }{\text{Cost MLMC} }$ \\[2mm]
\hline\hline \\[.01mm]
$0.0005$ & $0,1,2,3,4$ &$8.7\text{e}7$& $2.8\text{e}6$ & $2.1\text{e}6$  & $0.75$\\
$0.0005$ & $0,2,4$ & $6.6\text{e}7$ & $3.3\text{e}6$ & $1.9\text{e}6$  & $0.58$\\
$0.001$ & $0,1,2,3$ &$2.2\text{e}7$ & $6.5\text{e}5$ & $5.0\text{e}5$ & $0.77$ \\
$0.003$ & $0,1,2,3$ &$2.4\text{e}6$ & $7.3\text{e}4$ & $7.1\text{e}4$ & $0.97 $ \\
$0.005$ & $0,1,2,3$ &$8.7\text{e}5$ & $2.7\text{e}4$ & $3.7\text{e}4$ & $1.40 $ \\
\hline
\end{tabular}
\end{table}

\begin{center}
\begin{figure}
\resizebox{.5\textwidth}{!}{\includegraphics[trim= 10mm 55mm 10mm 60mm]{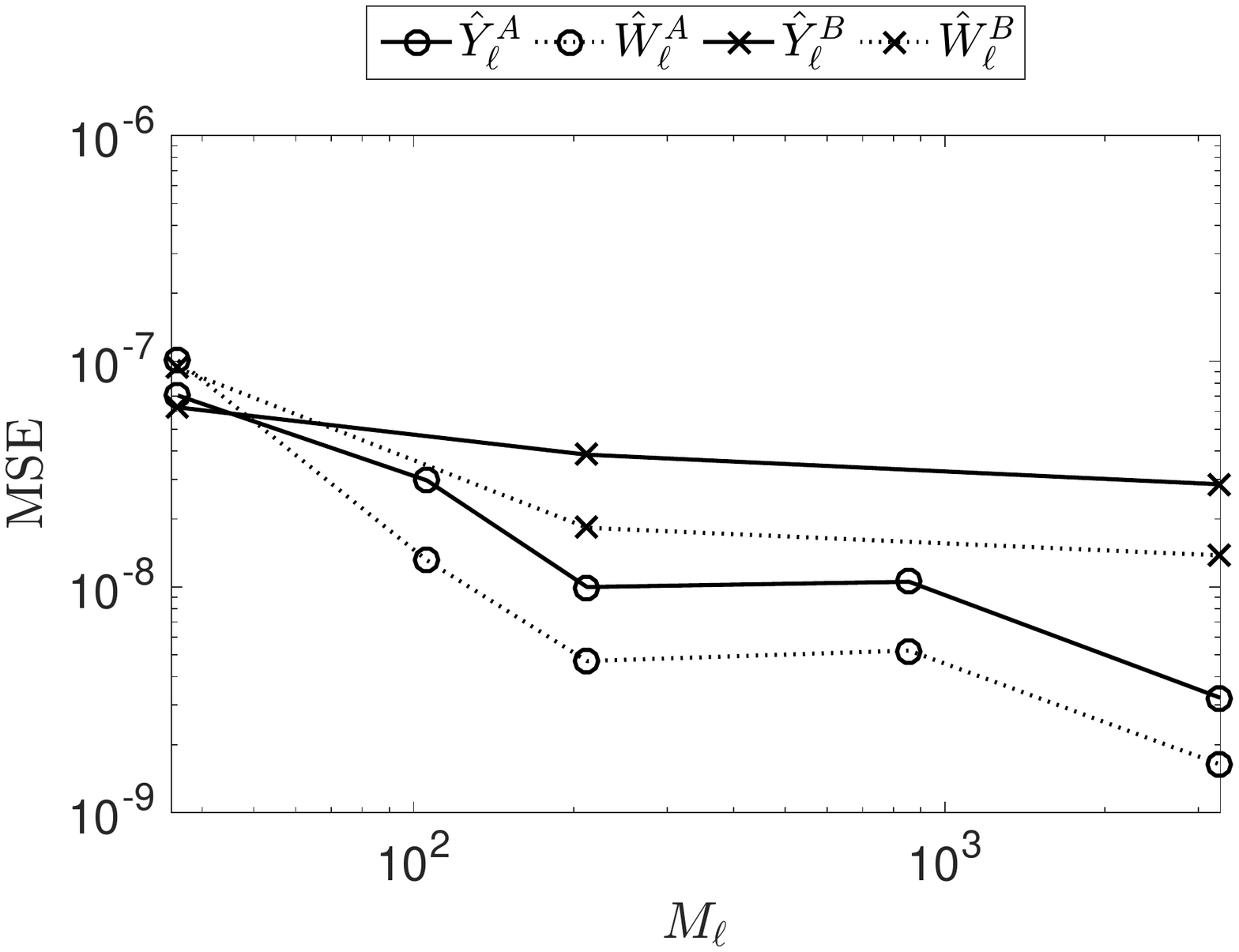}}
\resizebox{.5\textwidth}{!}{\includegraphics[trim= 10mm 55mm 10mm 60mm]{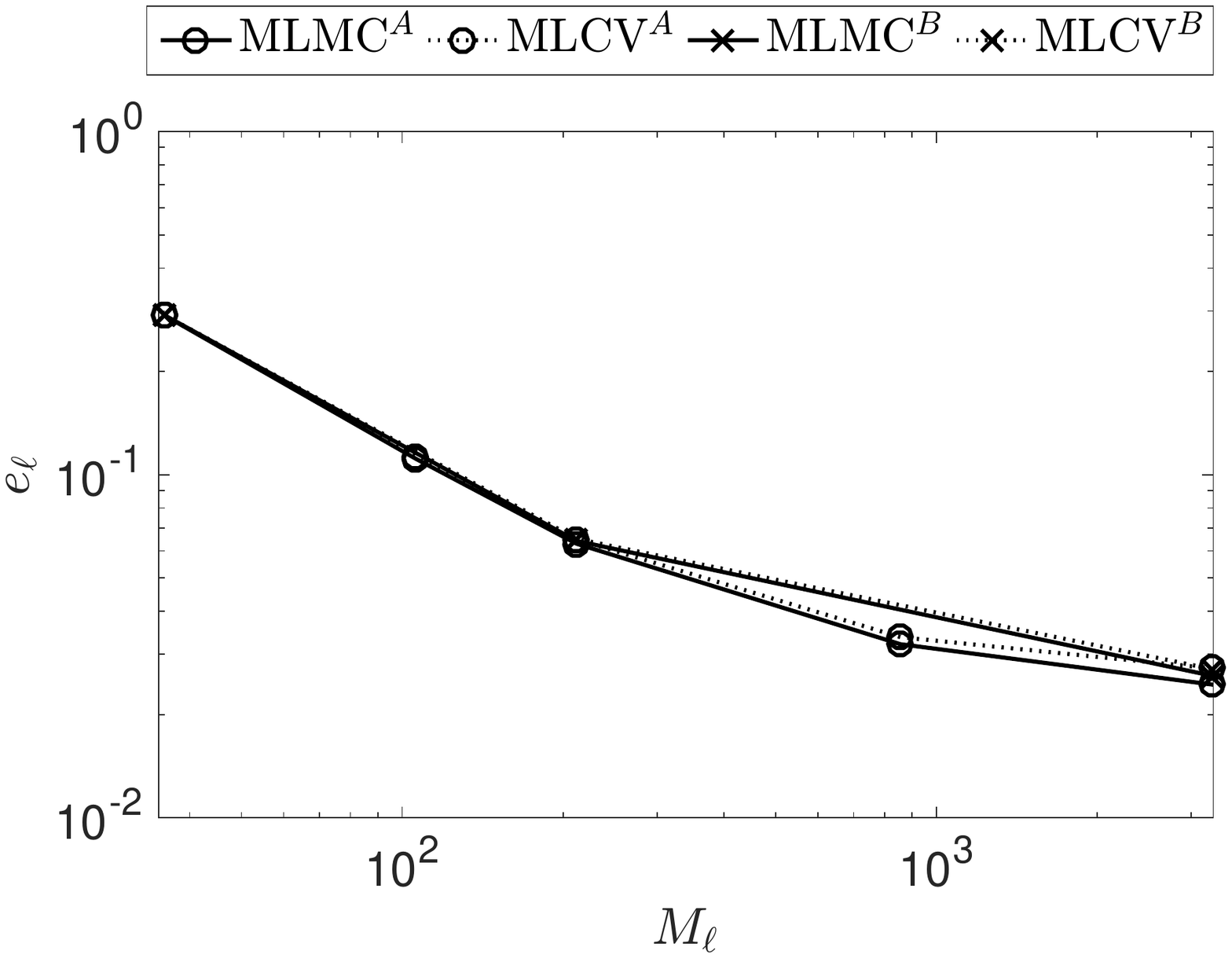}}
\captionof{figure}{(Left) MSE for $\hat{Y}_\ell$ and $\hat{W}_\ell$. (Right) Comparison of the relative errors (\ref{eq:rel_err_mlmc}) and (\ref{eq:rel_err_mlcv}) in estimating the mean QoI at each level for MLMC and MLCV. Superscripts $A$ and $B$ refer to the case $A$ and $B$ of the L-shaped domain problem, respectively.\label{fig:ERRL}} 
\end{figure}
\end{center}

These next results compare the estimated MSEs and relative errors of the methods. Figure \ref{fig:ERRL} (left) displays the estimated MSE for both methods. For both case {\it A} and {\it B}, the MSE estimates for $\hat{Y}_\ell$ and $\hat{W}_\ell$ are on the same order, which is expected. The slight difference is due to the number of terms in the sampling error. Since case {\it A} has more levels, most of the sampling error terms should be smaller than those of case {\it B}, as the total sampling error for both test cases must attain the same tolerance of $\varepsilon ^2/2$. 

To see how this new MLCV method performs against MLMC in terms of relative error, we compare the values computed in (\ref{eq:rel_err_mlmc}) and (\ref{eq:rel_err_mlcv}). Figure \ref{fig:ERRL} (right) displays the convergence of the relative error for both methods as the levels are refined. {\color{black}We note that, due to the cost of simulating the fine model, these results are based on an average of 10 runs of MLMC and MLCV, where the data in each run is not completely independent from data in other runs.} This result indicates that, experimentally, both methods converge on the same order. However the two methods do not cost the same to derive this result. As stated earlier, using the cost and the total number of samples on each level, it is found that to achieve a sampling error of $\varepsilon = 0.0005$, MLCV only requires about $75\%$ and $58\%$ of the computational cost of MLMC for case {\it A} and {\it B}, respectively.

\subsection{Test 2: Thermally Driven Cavity Flow}

\begin{figure}
\centering
\includegraphics[width =3.2 in]{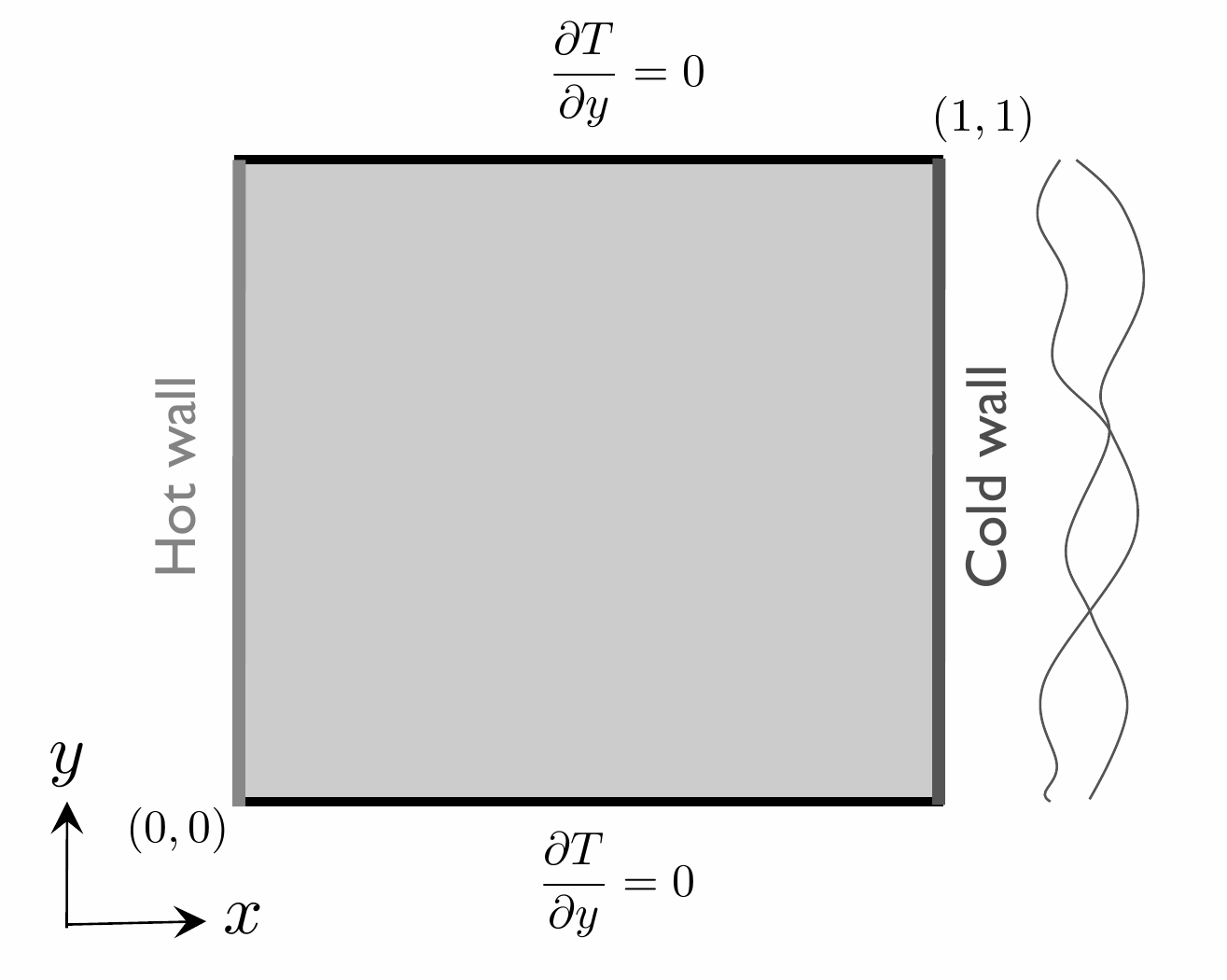}
\caption{Schematic of Test $2$, the thermally driven flow problem with random temperature along the hot (west) wall and spatially varying stochastic temperature along the cold (east) wall. \label{UC}}
\end{figure}

For the second implementation of MLCV, we consider a thermally driven flow problem in a square domain as described in \cite{LeQuere91,LeMaitre02b} and illustrated in Figure \ref{UC}. The left vertical wall has a random temperature $T_h$ with mean $\bar{T}_h$, while the right vertical wall, referred to as the cold wall, has a spatially varying stochastic temperature $T_c<T_h$ with constant mean $\bar{T}_c$. Both top and bottom walls are assumed to be adiabatic. The reference temperature and the reference temperature difference are defined as $T_{ref}=\bar{T}_c$ and $\Delta T_{ref}=T_h-\bar{T}_c$, respectively. Under small temperature difference assumption, i.e., Boussinesq approximation, the normalized governing equations are given by, \cite{LeMaitre02b},
\begin{equation}
\label{eqn:cavity}
\begin{aligned}
&\frac{\partial \bm{u}}{\partial t} + \bm{u}\cdot\nabla\bm{u}=-\nabla p + \frac{\text{Pr}}{\sqrt{\text{Ra}}}\nabla^2\bm{u}+\text{Pr}\Theta\bm{e}_y,\\
& \nabla\cdot\bm{u}=0,\\ 
&\frac{\partial \Theta}{\partial t}+ \nabla\cdot(\bm{u}\Theta)=\frac{1}{\sqrt{\text{Ra}}}\nabla^2\Theta,
\end{aligned}
\end{equation}
where $\bm{e}_y$ is the unit vector $(0,1)$, $\bm{u}=(u,v)$ is velocity vector field, $\Theta=(T- T_{ref})/\Delta T_{ref}$ is normalized temperature, $p$ is pressure, and $t$ is time. Zero velocity boundary conditions on all walls (in both directions) are assumed. For more details on the normalization of the variables in (\ref{eqn:cavity}), we refer the interested reader to \cite{LeQuere91,LeMaitre02b}. Prandtl and Rayleigh numbers are defined, respectively, as $\text{Pr}=\nu/ \kappa$ and $\text{Ra}={g}\tau\Delta T_{ref}{L}^3/({\nu}{\alpha})$. Specifically, $L$ is the length of the cavity, ${g}$ is gravitational acceleration, $\nu$ is kinematic viscosity, $\kappa$ is thermal diffusivity, and the coefficient of thermal expansion is given by $\tau$. In this example, we set $g=10$, $L=1$, $\tau = 0.5$, and $\text{Pr}=0.71$. We use finite volume for the discretization of (\ref{eqn:cavity}). The QoI $Q$ is the spatial average of steady-state heat flux along the hot wall. 

The uncertain quantities include the initial temperatures on the cold and hot wall ($T_c$ and $T_h$, respectively), as well as the viscosity parameter $\nu$. The cold wall temperature, $T_c$, is given by a Karhunen-Lo\`{e}ve-type expansion with $50$ terms, 
\begin{equation}\label{rbc}
T_c( y, {\bm \xi}) = \bar{T}_c + \sigma \sum_{i=1}^{50}\sqrt{\lambda _i} \phi _i (y) \xi _i,
\end{equation}

where $\lambda _i$ and $\phi _i(y)$ are the eigenvalues and eigenfunctions of the exponential correlation function
\begin{equation}\label{corr}
\mathcal{K}( y_1, y _2) = \sigma ^2 \exp\left(-\frac{\vert y_1 - y_2\vert}{l}\right),
\end{equation}
with $\sigma = 2$ and $l=0.15$. In (\ref{rbc}), we set $\bar{T}_c = 100$ and assume the random variables $\xi _i\sim U[-1,1]$, $i=1,\dots,50$, are independently and uniformly distributed over $[-1,1]$. Additionally, we assume $T_h\sim U[105,109]$ and $\nu\sim U[0.004,0.01]$. This brings the total number of random inputs to $d=52$. We note that the COV for the QoI of each simulation with this setup is about $25\%$. 

For the numerical setup of this problem, we consider four nested uniform grids with the sizes of $16  \times 16$, $32  \times 32$, $64  \times 64$, and $128 \times 128$. This implies we have $M_\ell = s^\ell M_0$, with $s = 4$. The resulting vector of heat flux values along the hot wall, $\q$, is of length $m_\ell = \sqrt{M_\ell}$. As stated earlier, since $m_\ell \ll M_\ell$, the added cost of performing ID and least squares is discounted. We select the rank, or equivalently, the size of the representative basis for $Q^{ID}_\ell$, to be $r =10$ for all levels. 

\subsubsection{Results of Test 2}

\begin{center}
\begin{figure}
\resizebox{.5\textwidth}{!}{\includegraphics[trim= 10mm 55mm 10mm 60mm]{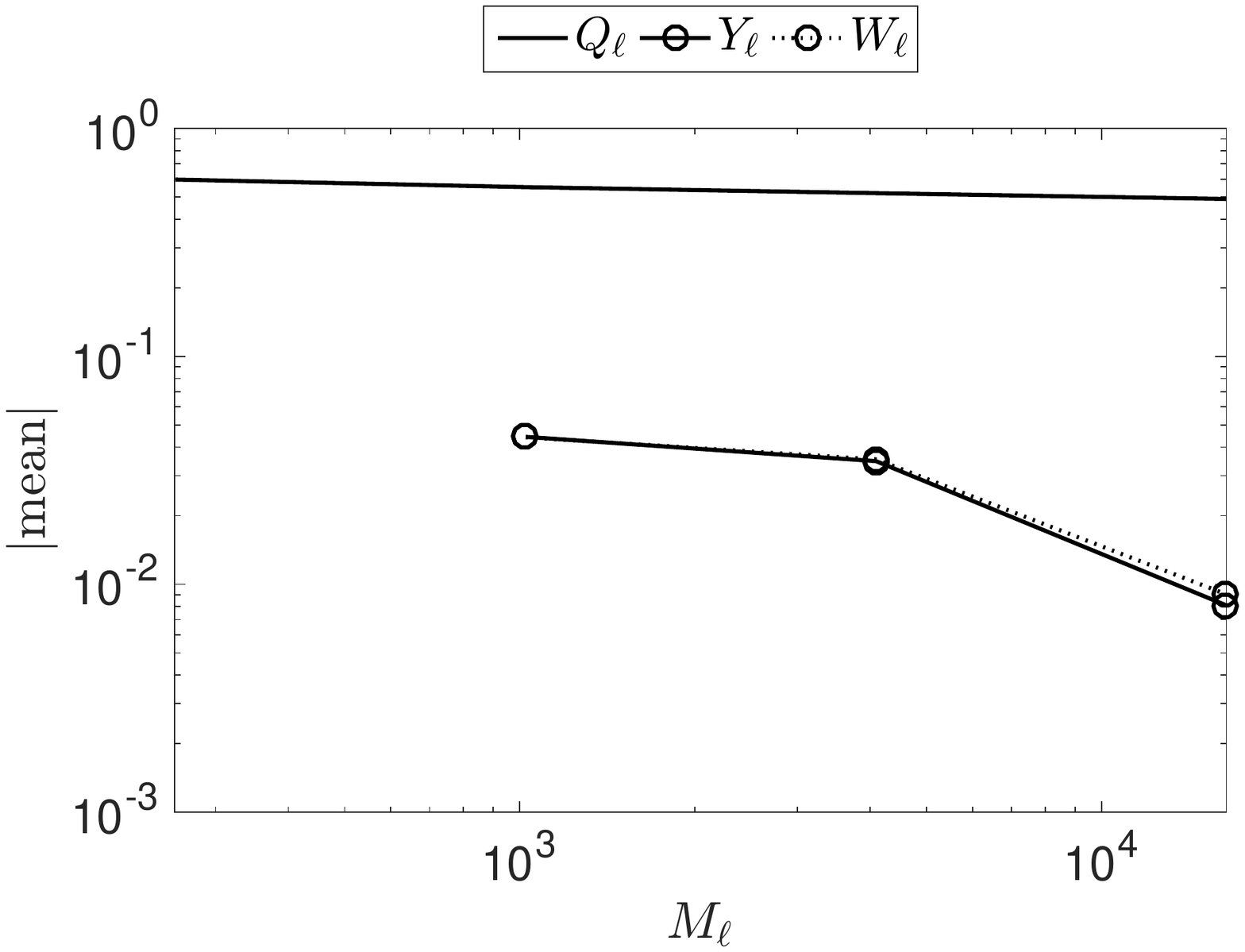}}
\resizebox{.5\textwidth}{!}{\includegraphics[trim= 10mm 55mm 10mm 60mm]{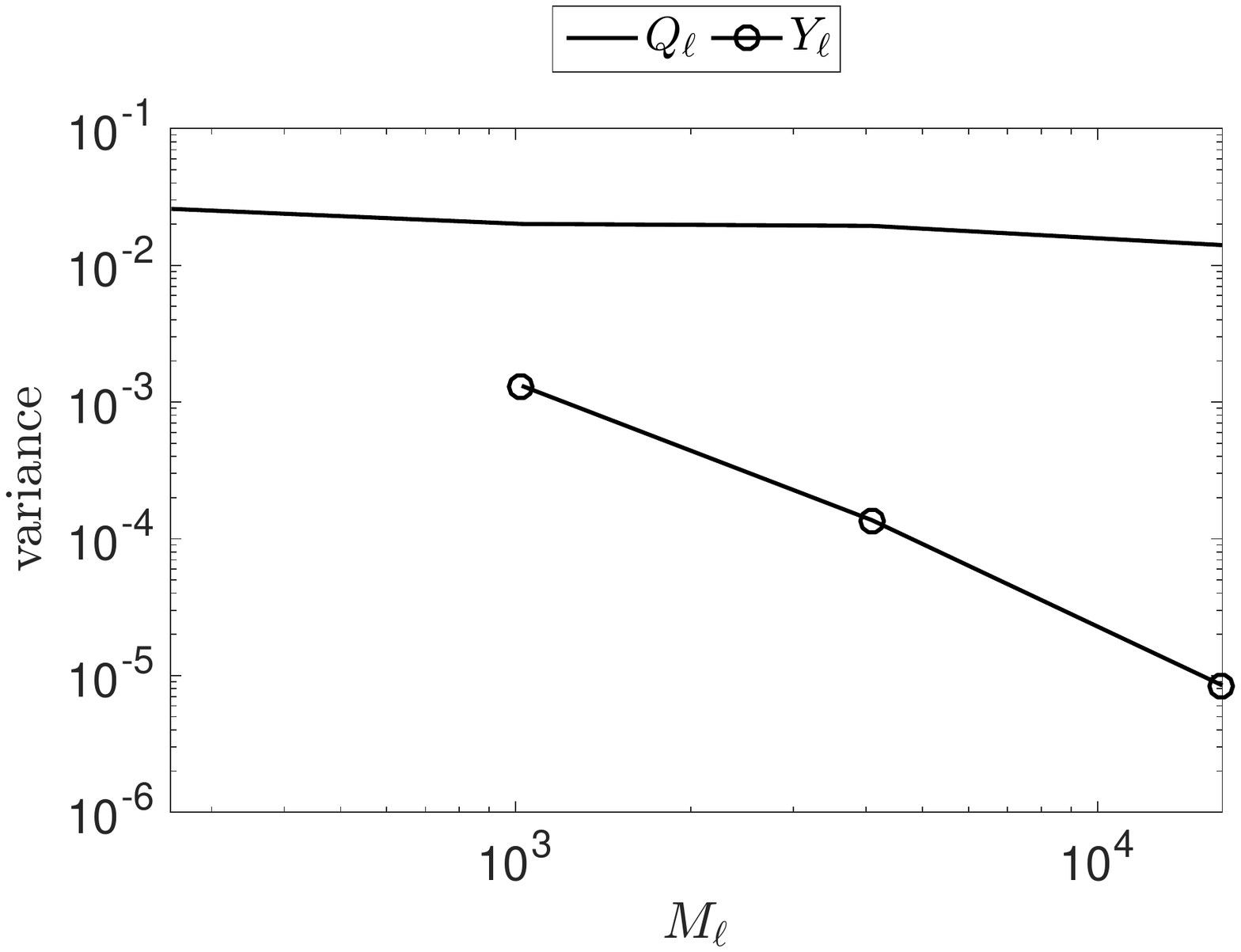}}
\captionof{figure}{(Left) Convergence of the sample mean of $Y_\ell$ and $W_\ell$, while that of $Q_\ell$ remains relatively constant. (Right) Convergence of the sample variance of $Y_\ell$ in comparison to that of $Q_\ell$. \label{fig:EVNL}}
\end{figure}
\end{center}

We first consider the results corresponding to the MLMC convergence guarantee in Section \ref{subsec:mlmc_theory}. Figure \ref{fig:EVNL} (left) shows the sample mean of $Q_\ell$, $Y_\ell$ and $W_\ell$ as a function of the degrees of freedom on each level, $M_\ell$. We note that the sample mean of $Q_\ell$ remains relatively constant, while those of $Y_\ell$ and $W_\ell$ decay at the same rate. The data displayed allows us to estimate the values of $\alpha$. Since the decay of data $Y_\ell$, $\ell=1,2,$ and $3$ does not show a clear convergence{\color{black}, i.e., $\ell=0$ discretization is {\it too coarse},} we use the data based on $Y_2$ and $Y_3$ to approximate $\alpha$. By doing so, we estimate $\mathbb{E}[Q_\ell-Q_{\ell-1}]\approx M_\ell^{-1.0}$, indicating that $\alpha \approx 1.0$. We note this estimate was then confirmed by using data obtained from the $256\times256$ mesh. Figure \ref{fig:EVNL} (right) shows the sample variance of $Q_\ell$ and $Y_\ell$ as a function of the degrees of freedom on each level, $M_\ell$. The data displayed allows us to estimate the values of $\beta$. We estimate $\mathbb{V}[Q_\ell - Q_{\ell-1}]\approx M_\ell^{-1.8}$, indicating that $\beta \approx 1.8$. By using the average CPU time to determine the cost of the solver, we find that $\gamma \approx 1.65$. This implies that the minimum MSE bound we can have is $\varepsilon \sim \mathcal{O}(0.00005)$ when using $M_L = 128^2$. Due to cost {\color{black}constraints}, we select $\varepsilon = 0.001$ for the {\color{black}numerical experiments}.

\begin{center}
\begin{figure}
\resizebox{.5\textwidth}{!}{\includegraphics[trim= 10mm 55mm 10mm 60mm]{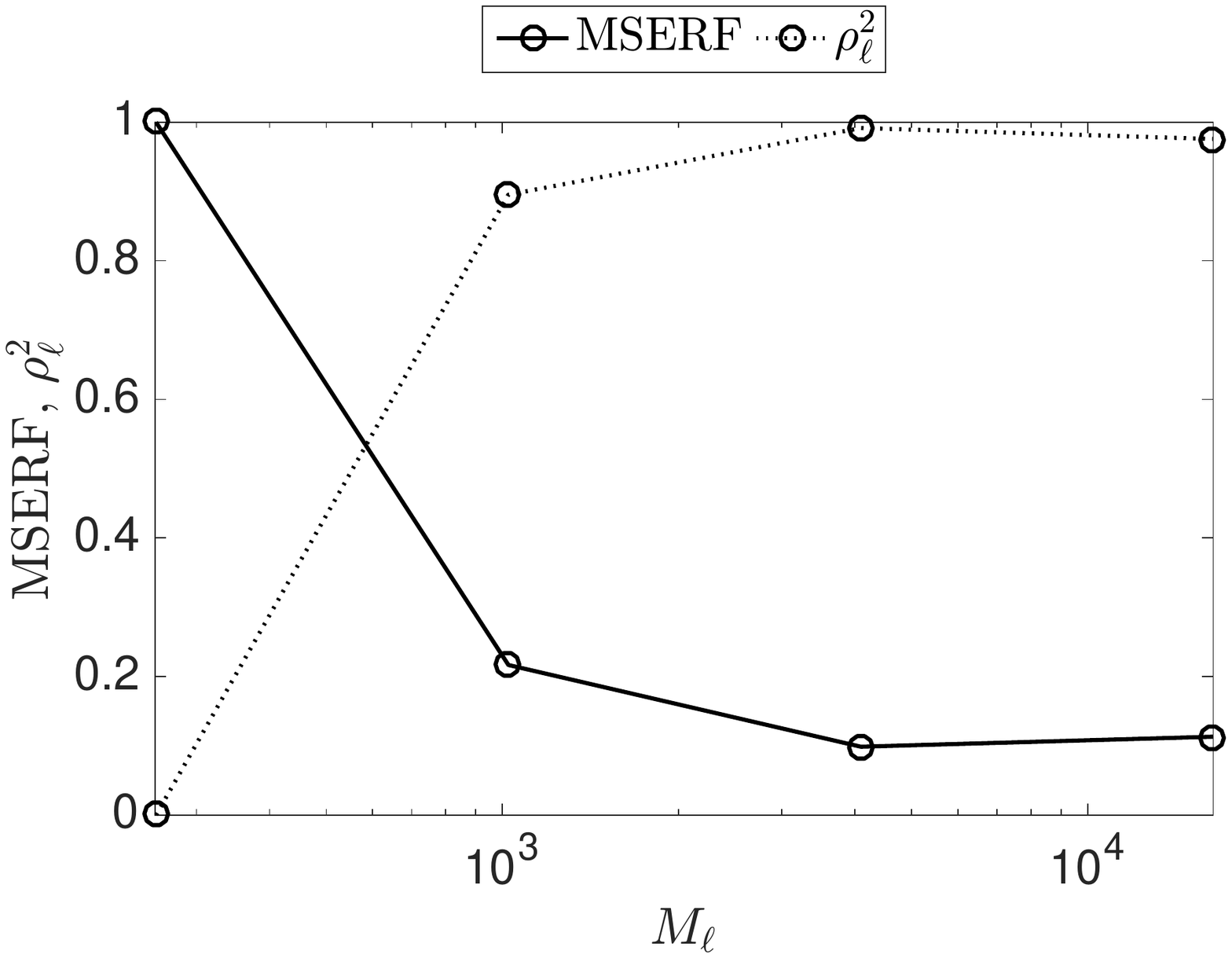}}
\resizebox{.5\textwidth}{!}{\includegraphics[trim= 10mm 55mm 10mm 60mm]{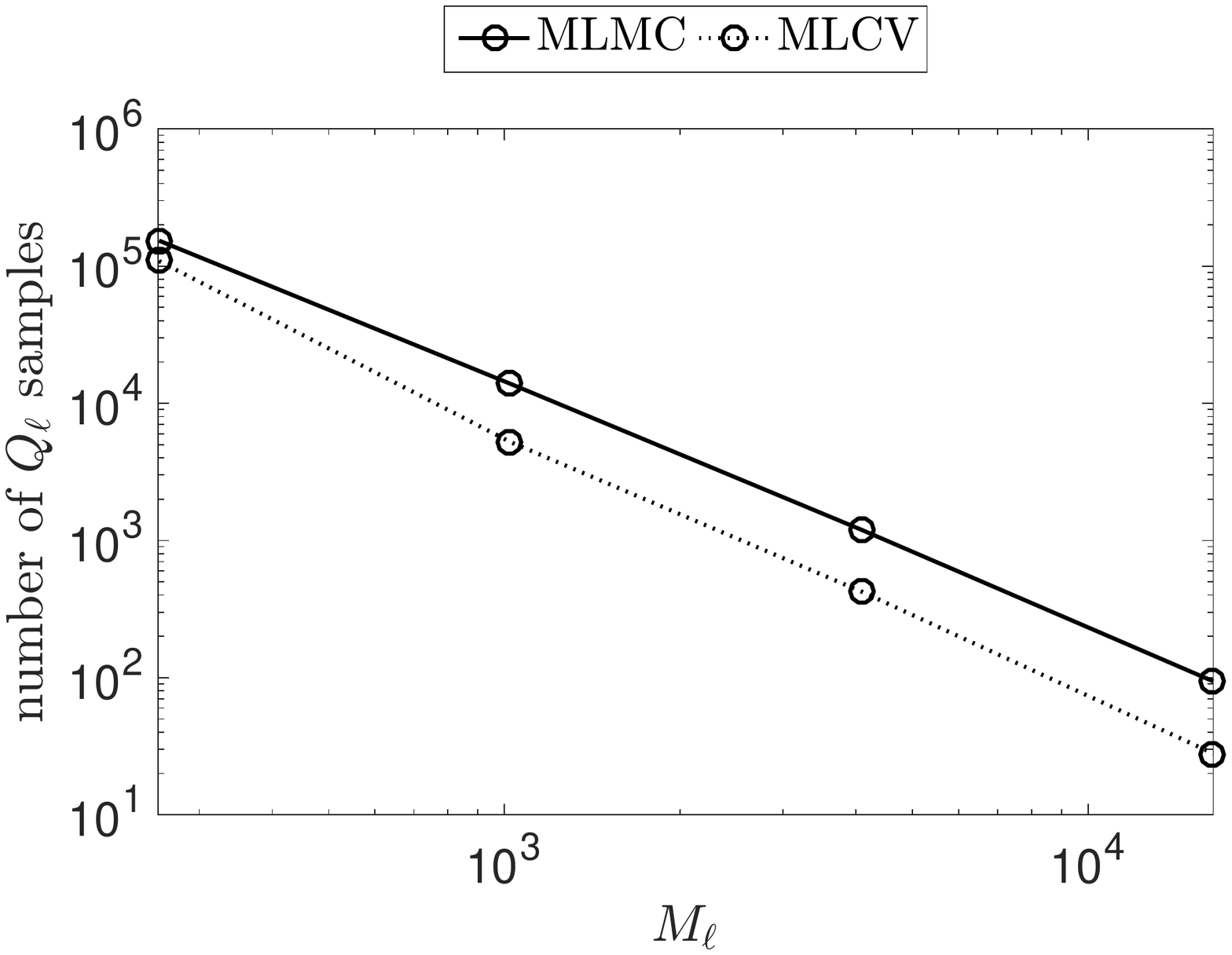}}
\captionof{figure}{(Left) MSE reduction factor (MSERF) between $\hat{W}_\ell$ and $\hat{Y}_\ell$, as well as the value of $\rho_\ell^2$ on each level. (Right) Total number of samples needed to attain the sampling error for MLMC and MLCV when $\varepsilon =0.001$. MLCV sample size includes the number required to determine $\bar{Z}_{\ell}$.\label{fig:MSENL}}
\end{figure}
\end{center}

Next we consider the MSE reduction as well as the sample size reduction in Figure \ref{fig:MSENL}. Figure \ref{fig:MSENL} (left) shows the values of $\rho_\ell^2$, where $\rho_\ell^2>0.85$ for levels $1,2$, and $3$, as well as the decay of the MSE reduction factor. As the MSE reduction factor approaches zero, the number of samples MLCV requires on each level decays. Figure \ref{fig:MSENL} (right) displays the number of samples required on each level for both methods when using $\varepsilon = 0.001$. It is clear that fewer samples are needed on each level for MLCV than for MLMC. And, as finer levels are approached, the gap between these two values increases. Because of this, we will see a significant cost payoff.

To compare the cost of performing MLMC and MLCV for this application, we consider the cost estimates displayed in Table \ref{t:2}, as determined by (\ref{eq:mlmc_totalcost}) and (\ref{eq:cost_mlcv}). Due to high simulation costs, we consider only four levels. We see that for $\varepsilon \leq 0.001 $ MLCV halves the cost of MLMC. As the value of $\varepsilon$ is increased, the cost difference of the two methods is smaller. For $\varepsilon \geq 0.005$, we can observe that MLCV has a larger computational cost than MLMC. The cost of the $r$ samples for the basis is diminishing the success of the MLCV method. 

\begin{center}
\begin{figure}
\resizebox{.5\textwidth}{!}{\includegraphics[trim= 10mm 55mm 10mm 60mm]{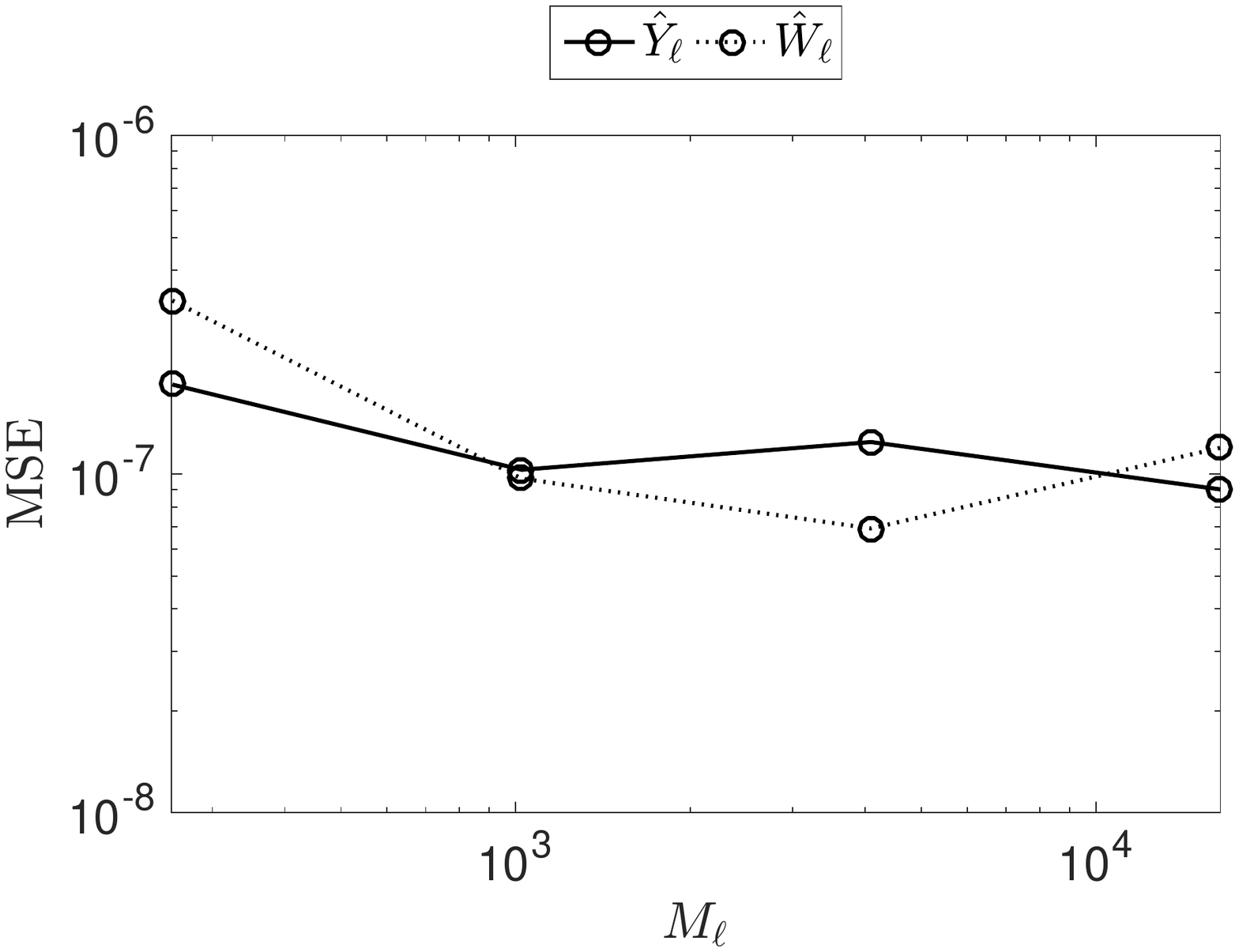}}
\resizebox{.5\textwidth}{!}{\includegraphics[trim= 10mm 55mm 10mm 60mm]{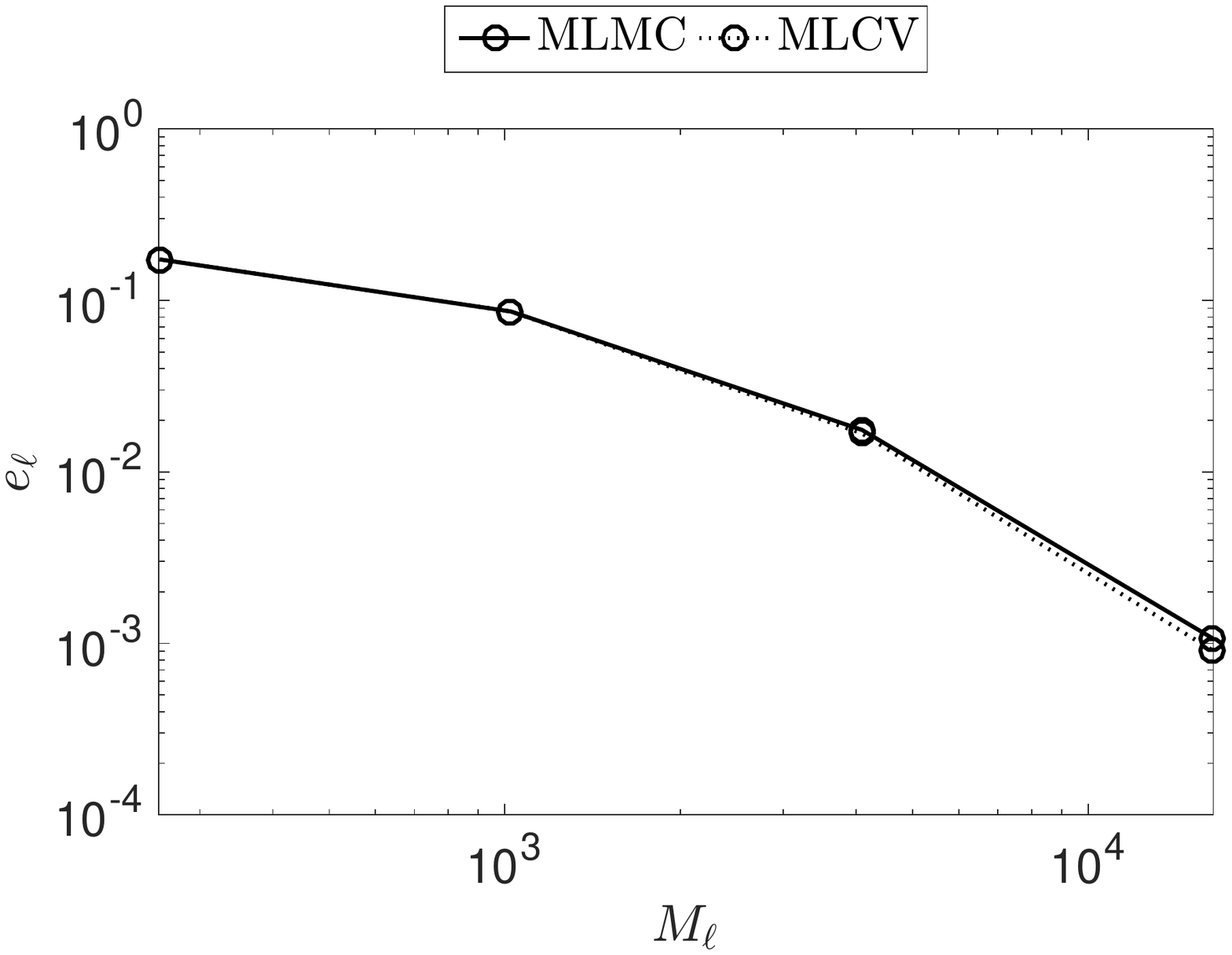}}
\captionof{figure}{(Left) MSE for $\hat{Y}_\ell$ and $\hat{W}_\ell$. (Right) Comparison of the relative errors (\ref{eq:rel_err_mlmc}) and (\ref{eq:rel_err_mlcv}) in estimating the mean of QoI at each level for MLMC and MLCV. \label{fig:ERRNL}}
\end{figure}
\end{center}

The final set of results compare the estimated MSEs and relative errors of the two methods. Figure \ref{fig:ERRNL} (left) displays the estimated MSE for both methods. We see that the MSE estimates for $\hat{Y}_\ell$ and $\hat{W}_\ell$ are on the same order. To determine the relative accuracy of the two methods, we next compute the relative errors (\ref{eq:rel_err_mlmc}) and (\ref{eq:rel_err_mlcv}). Figure \ref{fig:ERRNL} (right) displays the convergence of the relative error for both methods as the levels are refined. {\color{black}We note that, due to the cost of simulating the fine model, these results are based on an average of 10 runs of MLMC and MLCV, where the data in each run is not completely independent from the other runs.} As expected, we observe that as the mesh is refined for both methods, the relative errors improve. The notable result is that the relative errors for both methods remain essentially the same, while, in fact, MLCV only requires $48\%$ of the computational cost of MLMC.


\begin{table}[h]
\centering
\caption{Cost of MC, MLMC, and MLCV for Test 2 problem.\label{t:2}}
\begin{tabular}{llllll}
$\varepsilon$ & Levels &Cost MC&  Cost MLMC&Cost MLCV & $\frac{\text{Cost MLCV} }{\text{Cost MLMC} }$ \\[2mm]
\hline\hline \\[.01mm]
$0.00005$&$0,1,2,3$&2.2\text{e}9&$3.7\text{e}7$&$1.7\text{e}7$  & $0.45$\\[1mm]
$0.0005$&$0,1,2,3$&2.2\text{e}7&$3.7\text{e}5$&$1.7\text{e}5$  & $0.45$\\[1mm]
$0.001$&$0,1,2,3$&$5.6\text{e}6$&$9.3\text{e}4$&$4.4\text{e}4$  & $0.48$\\[1mm]
$0.003$&$0,1,2,3$&$6.2\text{e}5$&$1.0\text{e}4$&$6.8\text{e}3$  & $0.65$\\[1mm]
$0.005$&$0,1,2,3$&$2.2\text{e}5$&$3.8\text{e}3$&$4.0\text{e}3$   & $1.1$\\[1mm]
\hline
\end{tabular}
\end{table}

\subsection{Analysis}

The preceding results show a cost improvement over MLMC when applying this MLCV method. In addition, other observations regarding these results are important to highlight. In Test 1, the difference between the number of degrees of freedom on adjacent levels has a greater negative impact on the MLMC results, than those of MLCV. When using MLCV, the process does not need to be broken up into as many levels with small changes in the degrees of freedom as in MLMC, as MLCV does not rely solely on the difference between QoIs on adjacent levels. This method is especially beneficial in cases where there is {\color{black}not sufficient flexibility in defining multiple coarse spatial discretizations}, or when the solution converges slowly over the refinements. 

The cost ratio between MLCV and MLMC is also significantly different when comparing Test 1 and Test 2. While the $\rho_\ell ^2$ results for both methods are favorable, the faster cost growth between levels in Test 2 results in greater cost savings. More precisely, in Test 1 we had $\gamma =1$ and in Test 2 we had $\gamma = 1.65$. The payoff of placing more computational burden on the coarse level data than that of the fine level is more exaggerated for larger values of $\gamma$. This also suggests that MLCV will be even more advantageous over MLMC for three-dimensional (in physical space) problems, where $M_\ell$ grows faster.     

\section{Conclusion}
\label{sec:conclusion}

In this paper we have developed a non-intrusive, multilevel Monte Carlo method to approach uncertainty quantification problems with a large number of random inputs. The mathematical framework of this method relies heavily on the decomposition of the coarse level QoI data into a low-rank representation, from which a relatively accurate fine grid estimator can be calculated. Data with fast decaying singular values, combined with the use of control variates to minimize the MSE of the new variable results in this new MLCV method, with a total cost that is smaller than or comparable with that of MLMC. 

Both MLCV and MLMC methods are applied to a generalized eigenvalue problem associated with a linear elasticity model as well as a thermally driven cavity flow problem in a square domain. While the data in both tests lends itself to approximations with different ranks, and costs to determine the QoIs, MLCV was found to outperform MLMC. To reach a fixed MSE tolerance, MLCV was determined to require fewer samples, and thus a cheaper implementation than that of MLMC. 

\section*{Acknowledgements}

This work was supported by the United States Department of Energy under the Predictive Science Academic Alliance Program 2 (PSAAP2) at Stanford University and NSF grant CMMI-145460. This material is based upon work supported by the U.S. Department of Energy Office of Science, Office of Advanced Scientific Computing Research, under Award Number DE-SC0006402.

\section*{Appendix A}
\label{sec:appendix}

The following lemma from \cite{Martinsson11} outlines a number of key properties of the ID representation of $\bm{U}_{\ell-1} $, as seen in (\ref{eq:CGmid}). For simplicity, we drop the subscripts of $\bm{U}$, $\bm{C}$, $\bm{U}^{c}$, $m$, and $N$.

\begin{lemma} (Lemma 3.1 of \cite{Martinsson11}.) For any positive integer $r$ with $r\le\min\{m,N\}$, there exist a real $r \times N$  matrix  $\bm{C} $, and a real $m\times r$ matrix $\bm{U} ^c$ whose columns constitute a subset of the columns of $\bm{U} $, such that
\begin{enumerate}
\item some subset of the columns of $\bm{C} $ makes up the $r\times r$ identity matrix,
\item no entry of $\bm{C} $ has an absolute value greater than $1$,
\item $\Vert \bm{C} \Vert\le \sqrt{r(N-r)+1}$,
\item the least (that is, the $r$th greatest) singular value of $\bm{C} $ is at least $1$,
\item $\bm{U} = \bm{U} ^c\bm{C} $, when $r=m$ or $r=N$, and
\item $\Vert \bm{U} - \bm{U} ^c\bm{C} \Vert\le\sqrt{r(N-r)+1}\ \sigma_{r+1}$ when $r<\min\{m,N\}$, where $\sigma_{r+1}$ is the $(r + 1)$st greatest singular value of $\bm{U} $.
\end{enumerate} 
\end{lemma}

The construction of matrix ID relies primarily on the rank-revealing QR factorization of $\bm{U} $ given by
\begin{equation}
\label{eqn:RR_QR}
\bm{U}  \bm{P} \approx \bm{Q}\ [\ \bm{R}_{11}\ |\ \bm{R}_{12}\ ],
\end{equation}
where $\bm{P}$ is a $N\times N$ permutation matrix, $\bm{Q}$ an $m\times r$ matrix with orthonormal columns, $\bm{R}_{11}$ an $m\times r$ upper triangular matrix, and $\bm{R}_{12}$ an $r\times(N-r)$ matrix. In practice, the rank $r$ is unknown and thus the pivoted Gram-Schmidt process involved in (\ref{eqn:RR_QR}) is continued until $\Vert \bm{U} - \bm{U} ^c\bm{C} \Vert\le\epsilon$ is achieved for a predefined accuracy $\epsilon$. Given (\ref{eqn:RR_QR}), an $r\times(N-r)$ matrix $\bm{T}$ is sought for such that
\begin{equation}
\label{eqn:T_mat}
\bm{R}_{11} \bm{T} = \bm{R}_{12}.
\end{equation}
When $\bm{R}_{11}$ is ill-conditioned, \cite{Cheng05} suggests a solution $\bm{T}$ with minimum $\Vert \bm T\Vert_{F}$. Using (\ref{eqn:T_mat}) in (\ref{eqn:RR_QR}) we arrive at 
\begin{equation}
\label{eqn:RR_QR_1}
\bm{U}  \bm{P} \approx \bm{Q}\bm{R}_{11}\ [\ \bm{I}\ |\ \bm{T}\ ], 
\end{equation}
or, equivalently,
\begin{equation}
\label{eqn:RR_QR_2}
\bm{U} \approx \bm{U} ^c\ [\ \bm{I}\ |\ \bm{T}\ ]\bm{P}^{T} =  \bm{U} ^c\bm{C} , 
\end{equation}
where $\bm{U} ^c$ contains the first $r$ columns of $\bm{U}  \bm{P}$ and $\bm{C} =[\ \bm{I}\ |\ \bm{T}\ ]\bm{P}^{T}$.

\clearpage
\bibliographystyle{elsarticle-num}
\bibliography{AD_bib_v1}
\end{document}